\documentclass[prx,aps,superscriptaddress,nofootinbib,twocolumn]{revtex4-1}
\usepackage{mathrsfs}
\usepackage{amsfonts}
\usepackage{amssymb}
\usepackage{amsmath}
\usepackage{graphicx}
\usepackage[usenames,dvipsnames]{color}
\usepackage[colorlinks=true,citecolor=blue,linkcolor=magenta]{hyperref}
\usepackage{ulem}
\usepackage{lmodern}
\usepackage{enumitem}

\usepackage{bbm}
\newcommand\idop{\mathbbm{1}}

\newcommand\herm{^\dagger}

\newcommand\ox{\otimes}

\newcommand\sur[1]{^{(#1)}}
\newcommand\tr{\mathrm{tr}}

\newcommand{\Tr}{\mathrm{Tr}}

\newcommand{\ket}[1]{\vert{ #1 }\rangle}
\newcommand{\bra}[1]{\langle{ #1 }\vert}

\newcommand{\braket}[2]{\langle #1 \vert #2 \rangle}

\begin{document}

\title{An integrity measure to benchmark quantum error correcting memories}

\author{Xiaosi Xu}
\affiliation{Department of Materials, University of Oxford, Parks Road, Oxford OX1 3PH, UK}
\author{Niel de Beaudrap}
\affiliation{Department of Computer Science, University of Oxford, Parks Road, Oxford OX1 3PH, UK}
\author{Joe O'Gorman}
\affiliation{Department of Materials, University of Oxford, Parks Road, Oxford OX1 3PH, UK}
\author{Simon C. Benjamin}
\affiliation{Department of Materials, University of Oxford, Parks Road, Oxford OX1 3PH, UK}

\begin{abstract}
Rapidly developing experiments across multiple platforms now aim to realise small quantum codes, and so demonstrate a memory within which a logical qubit can be protected from noise. There is a need to benchmark the achievements in these diverse systems, and to compare the inherent power of the codes they rely upon. We describe a recently-introduced performance measure called {\it integrity}, which relates to the probability that an ideal agent will successfully `guess' the state of a logical qubit after a period of storage in the memory. Integrity is straightforward to evaluate experimentally without  state tomography and it can be related to various established metrics such as the logical fidelity and the pseudo-threshold. We offer a set of experimental milestones that are steps towards demonstrating unconditionally superior encoded memories. Using intensive numerical simulations we compare memories based on the five-qubit code, the seven-qubit Steane code, and a nine-qubit code which is the smallest instance of a surface code; we assess both the simple and fault-tolerant implementations of each. While the `best' code upon which to base a memory does vary according to the nature and severity of the noise, nevertheless certain trends emerge.
\end{abstract}

\maketitle

\section{Introduction}
Large scale quantum algorithms are expected to require hardware that is fault tolerant: small imperfections in the behaviour of physical qubits (whether they are superconducting loops, crystal defects or trapped ions) must be identified and corrected, so that there is no error on the logical level. Recently there has been rapid progress in the implementation of quantum codes, across platforms as diverse as ion traps~\cite{11Blatt, 14Blatt,17Monroe}, superconducting qubits~\cite{12Schoelkopf,15Martinis,15Chow,15diCarlo}, and crystal defect systems~\cite{16Taminiau}.

A comprehensively successful quantum code will have been achieved when one can demonstrate a full set of quantum operations on encoded qubits with a fidelity that exceeds that of the best possible unencoded physical qubits~\cite{gottesman2016quantum}. However this criterion is very challenging to achieve; it means `beating' the superb fidelities exceeding $99.9\%$ that can now be achieved with single physical qubits~\cite{16Wineland,16Ballance,2014Harty}. Even the task of achieving a superior coherence time with a memory based on an encoded qubit, versus a single physical qubit, is not trivial. Individually controlled physical qubits can persist for the order of a minute when not actively manipulated~\cite{2014Harty}, or $10$ minutes using dynamical decoupling~\cite{17Kim}. 

It is therefore interesting to find a measure for the efficacy of memories based on small quantum codes, using which we can identify reasonable milestones for near-future experimental realisations. Equally importantly we wish to be able to fairly compare memories based on platforms that might have very different inherent timescales. A number of measures of performance might be considered, including the diamond norm, the fidelity in the logical basis, surpassing the pseudo-threshold, and so forth.
Here we show that these measures can be related to a measure called the {\it integrity} of the logical qubit, which was recently introduced for assessing the performance of a memory based on the seven-qubit 2D color code (which is also the Steane code) in the context of ion trap quantum computing~\cite{17BermudezFeasibility}. 

Here we will motivate the notion of integrity through its intuitive meaning as ``the probability that Bob, receiving a logical qubit from the memory system, can infer its state''. We show that in simple cases integrity also corresponds to ``the fidelity of a logical qubit after storage in the memory'', but that the former meaning based on state inference remains meaningful even when the latter notion of a memory's fidelity becomes ill defined. We offer a set of four milestones based on comparing the integrity of an encoded and actively corrected quantum memory versus either un-corrected variant or with a single physical qubit. The milestones are increasingly challenging with the fourth being a demonstration of `Strictly superior encoded memory' . We report the results of a wide-ranging set of numerically intensive simulations, where we assess and compare several memories based on the five-qubit code, the seven-qubit Steane code, and the nine-qubit small surface code. We estimate the performance levels required in the error correcting process (performed by an agent we label `Igor') so that our milestones can be met. We establish that the task of evaluating the integrity of a memory is experimentally feasible when all the phases of the protocol (encoding, memory storage and decoding) are realised by the same imperfect hardware.

We conclude by discussing generalisations: it is straightforward and natural to extend the concept of integrity to encompass systems where a computation takes place. A further study of the properties of integrity appears in a partner paper to the present one~\cite{nielsPaper}. 

\section{Introducting integrity}
\label{sec:informal_Intro}
One can think of any memory as a channel for communicating information from the present ($t=0$) to a specified future time ($t=\tau$). The simplest notion of an ideal memory would be one where no change whatsoever happens to the stored information. Presently we will wish to generalise from this simple notion, but it is useful to begin by asking how we would benchmark performance against this basic standard: We could compare the state at $t=0$ with the state at $t=\tau$, using either the fidelity or the trace distance. Let us briefly review these two quantities.

There are two definitions of fidelity commonly used in the literature; one is the square of the other. Here we use the squared quantity, formally defining fidelity as
\begin{equation}
  \label{eqn:fidelity}
  \mathcal F(\rho_0,\rho_1) = \Bigl\lVert \sqrt{\rho_0} \sqrt{\rho_1} \Bigr\rVert_{\tr}^2.
\end{equation}
This definition uses the trace norm, itself defined as 
\begin{equation}
  \lVert \sigma \rVert_{\tr} = \Tr\Bigl(\sqrt{\sigma\herm \sigma}\Bigr),
  \label{eqn:traceNorm}
\end{equation}
and this is also the sum of the singular values of $\sigma$.

In the case that $\rho_0$ is a pure state $\ket{\psi_0}\bra{\psi_0}$, the fidelity then has a simple physical interpretation: if we measure state $\rho_1$ in a basis where one of the possible outcomes is $\ket{\psi_0}$, the fidelity is precisely the probability of this outcome. When both states are pure, we have simply
\[
  \mathcal F(\psi_0,\psi_1) = | \braket{\psi_0}{\psi_1} |^2.
\]

While the fidelity measures the similarity of two states, the trace distance measures the degree to which two states differ. It is defined as 
\begin{equation}
  \mathcal D(\rho_0,\rho_1) = \tfrac{1}{2}\bigl\lVert \rho_1 - \rho_0 \bigr\rVert_\tr.  
  \label{eqn:TD}
\end{equation}
Ranging from $0$ to $1$, the trace distance has a remarkably clear intuitive meaning: it tells us the probability that two states $\rho_0$ and $\rho_1$ could be told apart by an ideal experimentalist. Suppose that we present to an experimentalist, Bob, a theoretical description of both $\rho_0$ and $\rho_1$, and we also prepare a physical system in one of these two states (with a $50/50$ prior probability) and present this to the Bob. He must guess whether the physical state is $\rho_0$ or $\rho_1$. Using his optimal strategy, his probability $p_g$ of guessing correctly is simply
\begin{equation}
p_g=\tfrac{1}{2}+\tfrac{1}{2}  \mathcal D(\rho_0,\rho_1). 
\label{eqn:BobGuess}
\end{equation}
We will make extensive use of this idea presently. 

The functions $ \mathcal D(\rho_0,\rho_1)$ and $1- \mathcal F(\rho_0,\rho_1)$ can both be regarded as measures of how distinct two states are. However it is important to note that these quantities are fundamentally different, and can give very different `scores' in experimentally relevant cases. We opt to employ the trace distance, for various reasons described later but most particularly because Eqn.~\eqref{eqn:BobGuess} leads to straightforward experimental realisations. 

Our simple notion of an ideal memory -- one permitting no change -- is rather unsatisfactory. Certain changes are in fact harmless and do not practically reduce the quality of a memory.  Any deterministic, known and anticipated change to the stored information is harmless if we can easily compensate: an example is the continuous phase evolution which occurs within any physical qubit if the states $\ket{0}$ and $\ket{1}$ are non-degenerate eigenstates. For the present case of a memory that employs a quantum code to protect logical qubit(s) we can go further: Any {\it correctable} error is also relatively harmless in the sense that a ideal agent can recover the logical qubit with certainty. We would like our measure of the quality of a memory to incorporate these principles; additionally, we have a notion of a `useless' memory, one that should score zero, as a memory that fails to preserve any recoverable information whatsoever.

Suppose that some qubit with density matrix $\rho$ is to be stored in a code-based memory channel for a specified period of time. We will use the symbol $\Phi$ to denote the memory channel itself. The initial state $\rho$ maps to the final state $\tilde \rho$ through this process:
\begin{itemize}
\item {\bf Setup:} At $t=0$ Alice (taken to be perfect) encodes the single qubit $\rho$ into an $n$-qubit logical code: $\rho_n=\mathsf E(\rho)$ where $\mathsf E$ is the encoding map.
\item {\bf The memory channel:} Evolution and degradation of the logical qubit occurs while it is stored. This may include the effects of actively applied error correction cycles (involving a non-ideal agent, whom we label `Igor' and discuss presently).
  We have $\rho_n^\prime=\mathsf N(\rho_n)$ where $\mathsf N$ is the noise map.

\item {\bf Conclusion:} At $t=\tau$, Bob (taken to be perfect) performs an error correction cycle, and then reverses Alice's encoding process to obtain a single physical qubit: $\tilde\rho=\mathsf D(\rho^\prime_n)$ where $\mathsf D$ is the decoding map.
\end{itemize}

It is the second step that we are interested in; steps one and three (Alice and Bob) merely frame the process. We can write the entire channel as $\Phi = {\mathsf D \circ \mathsf N \circ \mathsf E}$, thus incorporating Alice's encoding $\mathsf E$, the noise $\mathsf N$, and Bob's decoding $\mathsf D$. This overall map $\Phi$ takes as input a single qubit state (Alice's initial choice $\rho$) and ultimately returns another single qubit state $\tilde\rho=\Phi(\rho)$, i.e. Bob's single qubit after decoding. 

For an initial concept of an ideal memory as one causing no change at all,
we would desire $\Phi(\rho)=\rho$, and so (for example) $1-\mathcal D(\rho,\Phi(\rho))$ could suffice as a good metric for the performance of our memory. However we have noted that a much broader notion of `ideal' is needed, for instance to accommodate systemic phase evolution.
Fortunately, there is a natural way to proceed: instead of focusing on the changes suffered by a single logical qubit between $t=0$ and $t=\tau$, we can instead focus on the idea that a memory should preserve the {\it distinguishability} of different states. This notion can incorporate both fixed, known evolutions and random-but-correctable errors. Conversely it will properly recognise that all forms of memory which leave us with no recoverable information, are equally and entirely useless.

Consider two pure states $\psi=\ket{\psi}\bra{\psi}$ and $\psi_\perp=\ket{\psi_\perp}\bra{\psi_\perp}$ which are orthogonal to one another. Orthogonal states have trace distance of unity, since they can certainly be told apart. Let Alice choose $\psi$ at random, uniformly from all possible single-qubit states, and then opt to encode {\it either} $\psi$ or instead the antipodal state $\psi_\perp$.  Then $\Phi(\psi)$ or $\Phi(\psi_\perp)$ will describe the state after it has passed through the memory channel and been decoded by Bob. If the channel has caused the same fixed evolution to occur to each (logical) state, or indeed if it has introduced errors but they are correctable, then these states will still be completely distinguishable -- they will still have trace distance equal to unity. Therefore we define the integrity of the memory as 
\begin{equation}
\boxed{
 \ \   \mathcal R(\Phi) =
    \min_{\psi}
      \mathcal D\bigl(\Phi(\psi), \Phi(\psi_\perp)\bigr).\ \ 
      \label{eqn:integrityShortDef}
}
\end{equation}
Note that we take the minimum over all possible choices of $\psi$ made by Alice. We do this to account for the fact that certain memory channels may have no detrimental effect on special choices of the state, as for example a dephasing channel leaves $\ket{0}$ and $\ket{1}$ unchanged.
To provide a measure which guarantees at least some quality of storage for {\it all} states, we consider the performance in the worst case.
Note that for many environmental noise models, including pure depolarising noise, Bob's performance does not vary with Alice's choice.

In the partner paper~\cite{nielsPaper} this definition is obtained from a more basic starting point where orthogonality is not imposed.
Our discussion proceeds from Eqn.~\eqref{eqn:integrityShortDef} for the sake of brevity.

It is worth emphasising that $R(\Phi)$ is a function on the memory channel $\Phi$ itself, thus one should speak of the {\it integrity of the memory} (including the specific choice of error correction technology).
It is understood that the memory channel is used for some defined time $\tau$, and that if the same memory system were used for a longer time then its integrity would be lower; typical channels will have zero integrity as $\tau\rightarrow \infty$. 

The integrity of the memory has a highly intuitive and natural meaning through the following scenario: We suppose that Bob initially knows nothing about Alice's choice of qubit to encode, but after Bob has completed his decode process to obtain the single qubit we then describe to him two choices: either Alice's initial qubit was $\psi$ or it was $\psi_\perp$. Bob then makes a measurement of his choice to try to determine whether it is $\Phi(\psi)$ or $\Phi(\psi_\perp)$ that he has received. The integrity $\mathcal R(\Phi)$ tells us Bob's probability $p_g$  of guessing successfully according to 
\begin{equation}
p_{g,{\rm worst}}=\tfrac{1}{2}+\tfrac{1}{2}  \mathcal R(\Phi)\ \ \ \text{and so}\ \ \mathcal R(\Phi)=2p_{g,{\rm worst}}-1.
\label{eqn:integrityRelatesToP}
\end{equation}
Here the label `worst' reminds us that this is the lowest success probability, i.e. when the options $\psi$, $\psi_\perp$ are the ones least well preserved by the memory (if indeed there is any variation). The integrity therefore describes the best possible guarantee that can be made on how well a memory preserves the distinctiveness of different states.

Notice that we constrain Bob to use a specific method to identify the received state. He must map the logical qubit back to a single physical qubit  by first applying a standard round of error correction for the code in question, then applying the inverse of Alice's encoding circuit. Bob's sole freedom is that he can choose how to measure that final physical qubit. As we explain in Appendix~\ref{uberBob}, by constraining Bob this way we ensure that his performance is associated with the code structure and its capacity to protect information.  In the Appendix we discuss the performance of a more powerful agent who is given full information about the error channel and complete license to perform any operations on all $n$ received qubits; this agent actually has very similar (sometimes identical) performance to our constrained Bob.

For some memories $\Phi$ (although not for all conceivable memories) Bob's correct strategy for his final step is the obvious one: just measure in the basis $\{\ket{\psi}, \ket{\psi_\perp}\}$ and make the guess correspondingly. Then Bob's success probability is simply the fidelity of the state $\Phi(\psi)$ with respect to Alice's initial state $\psi$, since the fidelity of any state with respect to a pure state is the probability of obtaining that outcome in a measurement  (as we remarked following Eqn.~(\ref{eqn:fidelity})). So in this case Bob's probability of guessing correctly is simply 
$p_g=\mathcal F(\psi,\Phi(\psi))$. Moreover his worst performance is 
\[
p_{g,{\rm worst}}=\min_{\psi}\mathcal F(\psi,\Phi(\psi)).
\]
But given Eqn.~\eqref{eqn:integrityRelatesToP} we can now offer a precise meaning to the idea of ``the (worst case) fidelity of a logical qubit stored in the memory'' for any channel where Bob would opt to measure in the basis $\{\ket{\psi}, \ket{\psi_\perp}\}$. For such a channel,
\[
F_{\rm logic}=\tfrac{1}{2}+\tfrac{1}{2}  \mathcal R(\Phi).
\]
Loosely, $F_{\rm logic}$ is the fidelity after we project into the logical subspace of the code with a perfect round of error correction.
For memory channels with sufficiently complex noise maps $\mathcal N$ that Bob's choice of measurement basis would {\it not} be $\{\ket{\psi}, \ket{\psi_\perp}\}$, the very idea of the ``fidelity of a logical qubit stored in the memory'' becomes ill defined. Thus, integrity is a general measure which relates to the notion of logical fidelity {\it when the latter notion makes sense}. However integrity remains well-defined and meaningful even when the logical fidelity does not: it is the more general and robust concept.

Importantly, it is eminently practical to directly measure integrity in an experimental setting. Notice that although the definition Eqn.~(\ref{eqn:integrityShortDef}) refers to two different states, we would evaluate the integrity $\mathcal R$ through a series of single uses of the memory -- we simply follow our scenario described above involving Bob guessing between options and employ Eqn.~\eqref{eqn:integrityRelatesToP}. Consequently the costly process of performing full state tomography is not required. Equally importantly, while the definition describes the encoding and decoding as occurring perfectly (conceptualised by saying that Alice and Bob are perfect), we will show that in practice they can be made imperfect and yet the experiment can gauge the integrity with good accuracy. These features are discussed in more detail in Section~\ref{sec:realExp} and Appendix~\ref{appendix:whenIsPauliEnough}.

\begin{table}
  \centering
  \begin{tabular}{|c|l|} \hline
    & Theoretical protocol for measuring integrity \\ \hline\hline
    1a&  Alice (perfect) prepares a single qubit state $\ket{\psi}$ or  $\ket{\psi_\perp}$.\\ 
    1b& Alice perfectly encodes it into the $n$ physical qubits.\\ \hline
    2 & From $t=0$ to $\tau$ the $n$ qubits are in the memory; noise\\
    &  occurs from environment and possibly error correction.\\ \hline
    3a & Bob (perfect) performs error correction on the $n$ qubits,\\ 
      &   then decodes (inverse of 1b) the state to a single qubit. \\ 
    3b & Bob is told Alice's qubit was {\it either} $\ket{\psi}$ {\it or} $\ket{\psi_\perp}$. He\\ 
     & measures his qubit and guesses, success probability $p_{g}$.\\ \hline             
  \end{tabular}
  \caption{  Evaluating the integrity of a memory system. See also Fig.~(\ref{Fig:IgorFig}).}
  \label{table:basic_protocol}
\end{table}

\section{Milestones toward successfully protected memories}
\label{milestones}

Armed with this notion of the integrity $\mathcal R$ of a memory channel $\Phi$, in essence the worst-case probability that the state of a stored qubit can be inferred by Bob, we now identify milestones towards the goal of superior code-based quantum memories.

For convenience of exposition we may imagine that a third party, besides Alice and Bob, is responsible for the cycle(s) of error correction performed during the memory period: since this individual is effectively a flawed assistant for Bob, we use the name Igor after the famous fictional lab assistant. 
We initially focus on the case where at most one error correction cycle is used during the entire period $\tau$ where memory operates, i.e. in between Alice ($t=0$) and Bob ($t=\tau$). We therefore now specify Step~2 of Table~\ref{table:basic_protocol} in more detail, setting it out in Table~\ref{table:first_criterion}. The key idea will be to compare the integrity of the memory channel without error correction (no Igor) to the case with EC (Igor participates) and determine whether the latter is superior. 

Let us use the symbol  $\Phi^m$ to label the memory channel when Igor performs $m$ rounds of error correction, so that  $\Phi^0$ labels the channel when no QEC is performed (i.e. noise sources are purely environmental). Then we say that a round of error correction is beneficial if Bob's probability of subsequently discriminating the state correctly is higher when Igor indeed performs that round, i.e. when 
\begin{equation}
 \mathcal R(\Phi^1) >  \mathcal R(\Phi^0). 
\label{eq:IgorIsGood1}
\end{equation}
This criterion for successful error correction can be summarised as, ``Is Igor a help or a hinderance to Bob?". 

This seems entirely straightforward but there is a subtlety: the question of whether Eqn.~(\ref{eq:IgorIsGood1}) is satisfied will depend on the duration $\tau$ of the memory channels (here we would naturally set the same $\tau$ for both $\Phi^1$ and $\Phi^0$ for a fair comparison). Since we are interested in defining a first milestone for experimental efforts, we say that error correction {\it can be} beneficial if
\begin{equation}
\mathcal R(\Phi^1) >  \mathcal R(\Phi^0)\ \ \ \text{for some value of }\tau.
\label{eq:IgorIsGood}
\end{equation}
We will refer to the challenge of satisfying Eqn.~(\ref{eq:IgorIsGood}) as milestone {\textbf{M1: Beneficial error correction}}\label{M1:BEC}.

One might wonder if we should consider a stronger condition: $\mathcal R(\Phi^1) >  \mathcal R(\Phi^0)$ for {\it all values} of the common duration $\tau$. It is interesting and important to note that this condition will be impossible to satisfy in physical devices where the process of error correction is very fast compared to the rate of environmental decoherence. This applies, for example, to ion trap devices with clock-transition qubits where the environmental decoherence time can be on the order of minutes but gate operations are sub-millisecond. We need only assume that environmental decoherence is a continuous process to see that $\mathcal R(\Phi^0)\rightarrow 1$ as $\tau\rightarrow 0$, i.e. the integrity of the simple memory channel is arbitrarily close to unity for a sufficiently short value of the memory duration $\tau$. In other words, finite environmental noise needs finite time to occur. We can inspect the equivalent limit for $\Phi^1$ if we  assume that Igor's error correction cycle is instantaneous (whereas if it takes finite time $\delta$ then we cannot employ memory channel $\Phi^1$ for time durations less than $\tau<\delta$). But given instantaneous error correction, we find $\mathcal R(\Phi^1)\rightarrow \epsilon$ as $\tau\rightarrow 0$, where $\epsilon$ is non-zero and is related to the inevitable imperfections in the cycle of error correction, i.e. the circuit elements such as state initialisations, one- and two-qubit gates, and measurements will all have finite infidelity. What we are noticing is that it is not desirable to perform error correction `as frequently as possible' -- we should wait for a finite time before applying an error correction cycle, so that its negative impact on the memory is justified by the positive gain. This is made very apparent by Fig.~\ref{fig:multi} in the next section. Of course, if the time required for error correction is comparable to the environmental decoherence rate, as may be the case for superconducting qubits, then one never has the luxury of waiting until the optimum time to perform error correction; cycles should indeed be performed `back to back'.

\begin{figure}
 \begin{centering}
  \includegraphics[width=1.02\columnwidth]{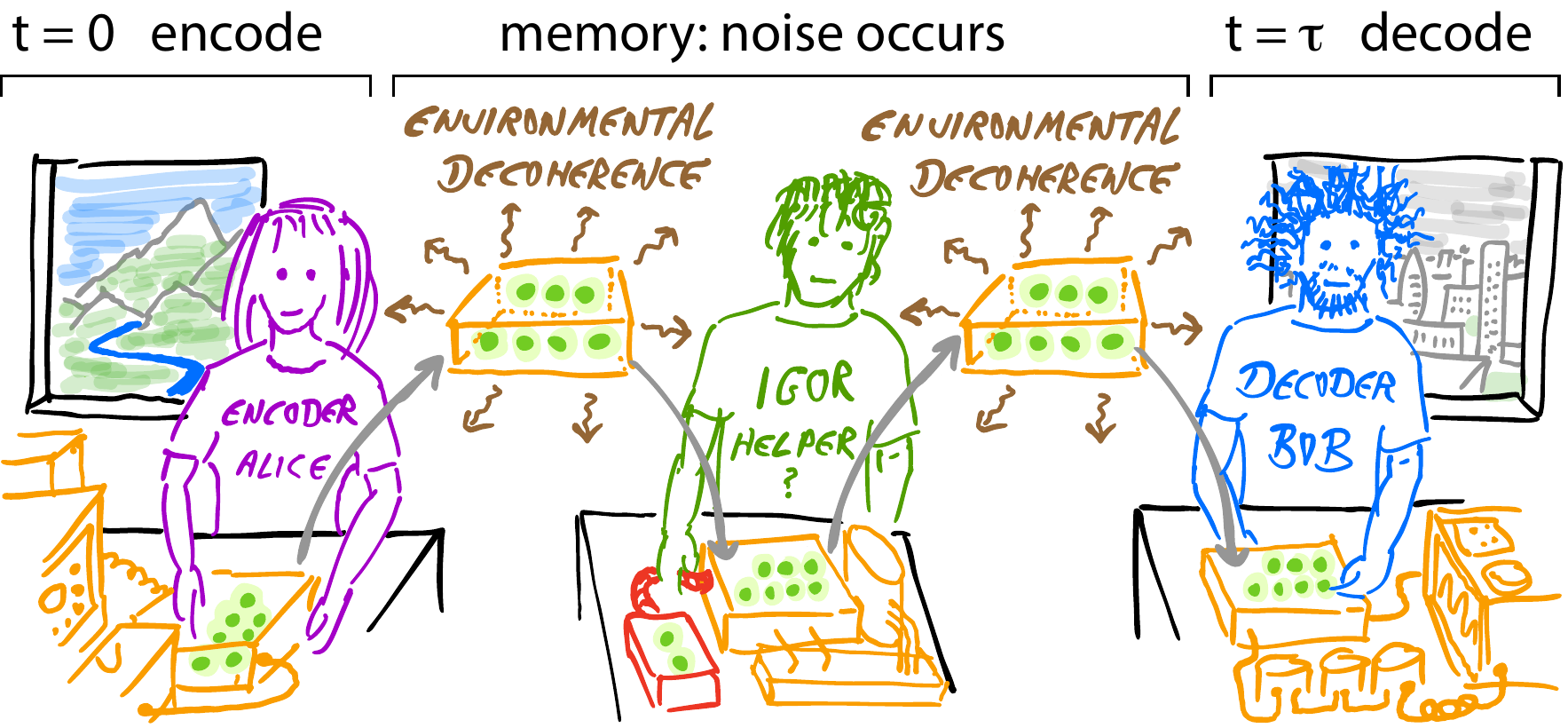}\\
  \caption{\label{Fig:IgorFig} {\bf Adapted from~\cite{17BermudezFeasibility}: Cartoon of the protocol for assessing the integrity of an error-corrected memory}. Alice and Bob perfectly perform the encoding and measurement, respectively, of a logical qubit. Meanwhile Igor is an imperfect agent attempting error correction to fight noise.}
\end{centering}
\end{figure}

\begin{table}
  \centering
  \begin{tabular}{|c|l|} \hline
    \ & {\bf Without Error Correction: Memory $\Phi^0$} \\ \hline
    2a & The $n$ physical qubits are subjected to\ \ \ \ \ \ \ \ \ \ \ \ \ \ \ \\ 
     & environmental noise for a time $\tau$.
\\ \hline             
  \end{tabular}
  \\ \  \\
  \begin{tabular}{|c|l|} \hline
    \ & {\bf With Error Correction: Memory $\Phi^1$} \\ \hline
    2a & The $n$ physical qubits are subjected to \\ 
     & environmental noise for a time $(\tau-\delta)/2$.\\ \hline
    2b & Optionally, Igor  is asked to apply a full round of \\
    & {\it imperfect} error correction, taking time $\delta$.\\ \hline
    2c& The $n$ physical qubits are subjected to \\
      & environmental noise for a further time $(\tau-\delta)/2$. 
\\ \hline             
  \end{tabular}
  \caption{Expanding on Step 2 of Table~\ref{table:basic_protocol} when we wish to assess the benefits of error correction.}
  \label{table:first_criterion}
\end{table}

While Eqn.~(\ref{eq:IgorIsGood}) is an important first milestone for an error-correcting quantum memory, further milestones can also be identified. For a sufficiently high performing Igor, and  a long memory duration $\tau$, it will be beneficial to have multiple rounds of correction.  This will be a signature of further progress toward a practical quantum memory. We would then find that 
\begin{equation}
 \mathcal R(\Phi^m) >  \mathcal R(\Phi^{m-1}) \ \ \ \text{for some value of }\tau.
\label{eq:nRounds}
\end{equation}
Here we understand this need only be satisfied for some particular $m>1$ (it seems likely that it would be achieved first for $m=2$ but we do not insist on this). We will refer to the challenge of meeting the condition in Eqn.~(\ref{eq:nRounds}) as milestone \textbf{{M2: Beneficial multi-round error correction}}\label{M2:BMREC}.

Equations~(\ref{eq:IgorIsGood}) and~(\ref{eq:nRounds}) involve comparisons between memories which both employ encoded qubits. There is of course another type of comparison we can make, one which directly addresses the question of whether it is `worth' using encoded memories at all: we should contrast such a memory channel to a simple, single qubit memory. Let us use the symbol $\Theta$ for that memory channel. We can consider its integrity $\mathcal R(\Theta)$ easily enough. Alice prepares a single qubit, again choosing between $\psi$ and $\psi_\perp$, but does not encode it into multiple qubits. It exists as a memory from $t=0$ to $t=\tau$, and finally Bob receives it but of course he has no decoding to do. The qubit he receives, $\Theta(\psi)$ or $\Theta(\psi_\perp)$, differs from Alice's qubit only because of environmental noise. But as before Bob must measure it to guess between the two possible states, and as before his probability of success is simply $\tfrac{1}{2}+\tfrac{1}{2}\mathcal R(\Theta)$. For our actively-corrected encoded memory to `beat' the simple single-qubit memory, we require
\begin{equation}
\label{eq:versusSingleQubitMem}
\mathcal{R}(\Phi^m)>\mathcal{R}(\Theta)\ \ \text{for {\bf some}}\ \tau_\Theta,\ \text{while using}\ \tau_\Phi=\alpha\tau_\Theta.
\end{equation}
Here we require only that this is true for some specific value of $m>0$ (it seems likely that $m=1$ would be the first demonstration). Note the more complex condition on the channel durations. In Eqn.~(\ref{eq:IgorIsGood}) and Eqn.~(\ref{eq:nRounds}) it was clear that the duration $\tau$ of the two memory channels should be the same for a fair comparison. This is not necessarily true of the Eqn.~(\ref{eq:versusSingleQubitMem}) since one can argue that the meaningful time scale for a quantum memory is not `wall clock' time but rather the time required to perform an active gate operation (perhaps the average time, given that circuit operations will differ in their time requirements, or perhaps the slowest time to be strict). Depending on the hardware platform and architecture, the time required to perform a gate operation on an encoded qubit may be longer than the time to perform the equivalent operation on an unencoded qubit. This would then suggest that $\tau_\Phi$ should be longer than $\tau_\Theta$, and the factor $\alpha\geq 1$ is included in Eqn.~(\ref{eq:versusSingleQubitMem}) to reflect this. We will refer to the challenge of satisfying Eqn.~(\ref{eq:versusSingleQubitMem}) as milestone \textbf{{M3: Beneficial encoded memory}\label{M3:BEM}}.

Here we can make contact with the concept of a `pseudo-threshold' (see e.g. Ref.~\cite{cross2009}). This concept is typically used in the context of concatenated codes, where there may be several levels of concatenation required before error rates fall sufficiently for deep quantum algorithms (such as Shor's or particularly Grover's algorithm). In the present context, we restrict our interest to the lowest level of concatenation where a process involving unencoded qubit(s) is compared to a process with a single level of encoding. The pseudo-threshold has been surpassed if a circuit performs to a higher standard with the encoded qubits, i.e. logical qubits, versus using the physical qubits directly. One might demand that for a complete universal set of operations, each operation at the encoded level outperforms the analogous operation using unencoded qubits. Alternatively one might speak of the pseudo-threshold for a specific operation, such as a single-qubit gate, a {\small CNOT} operation, a measurement or indeed a memory. In essence Eqn.~(\ref{eq:versusSingleQubitMem}) represents the (lowest tier) pseudo-threshold for memory, i.e. for the identity operation in a circuit.

Assuming that Eqn.~(\ref{eq:versusSingleQubitMem}) can be satisfied, there is a higher goal which might be achieved namely
\begin{equation}
\label{eq:unconditionallyBetter}
\max_m\mathcal{R}(\Phi^m_{\mathrm{QEC}})>\mathcal{R}(\Theta)\ \ \text{for {\bf all}}\ \tau_\Theta,\ \text{and using}\ \tau_\Phi=\alpha\tau_\Theta.
\end{equation}
Here the maximum is over a family of memory channels having the same duration $\tau_\Phi$ but with differing numbers of error correction cycles $m$. Importantly, we permit $m=0$. If this condition is satisfied, it means that for any desired duration we can sustain our encoded quantum memory at a higher integrity than a single physical qubit memory. We do so by applying a suitable number of error correction cycles. Moreover this is true even allowing for the factor $\alpha$ discussed above. This is therefore the `gold standard' for demonstrating a quantum memory and it is the most challenging of the criteria we have presented in this section. We will refer to the task of satisfying Eqn.~(\ref{eq:unconditionallyBetter}) as milestone \textbf{{M4: Strictly superior encoded memory}}\label{M4:SSEM}.

We have presented four milestones in an order which we expect may represent an increasing degree of challenge. It is not necessarily the case that each is more difficult than the last -- for example, conceivably M3 may be achieved before M2 in a given physical device. However the fourth milestone is clearly the most demanding and we should expect that the inequalities in Eqns.~(\ref{eq:IgorIsGood}), (\ref{eq:nRounds}) and (\ref{eq:versusSingleQubitMem}) must all be satisfied before Eqn.~(\ref{eq:unconditionallyBetter}) can be achievable.


\section{Numerical studies }
\label{numerations}
In this section we present our simulation results under the Alice-Igor-Bob framework described in Tables~\ref{table:basic_protocol} and \ref{table:first_criterion}. The simulation technique is based on the Monte Carlo method, the advantage of which has been described in~\cite{17BermudezFeasibility}. We emphasise that the integrity benchmark we have described is appropriate for any and all error models, including coherent noise, non-Markovian noise and so on. As specified in Appendix~\ref{appendix:errorModel}, in this paper we have used the simple canonical Pauli depolarising noise model (on all elements including state preparations, gates, measurements, and environmental noise) since it is a standard model to use in a first investigation. We aggregate a large number of individual runs, in each of which a pure state undergoes a specific trajectory: after every circuit element is applied, a classical random number is generated and compared with the error rate for that circuit element in order to decide whether an error is applied and if so its type. Each data point presented in this paper is a result of at least one million runs, and in order to make a smooth curve, at least 50 data points are generated for a single curve. The hardware used for this work is a cluster of approximately 100 nodes, which are connected by Intel TruScale QDR Infiniband. Each node is based on a motherboard with two Intel E5-2640v3 CPUs and has between 64 and 256GB of memory. 

For all the simulations presented in this paper, we use the Alice-Igor-Bob scenario that has been discussed in Tables~\ref{table:basic_protocol} and~\ref{table:first_criterion}. The circuit level description is shown in Figure~\ref{fig:5qubitScheme}, where we take the five-qubit code as an example. Igor performs his error correction cycle halfway through the duration of the memory channel (or for a channel with $n$ correction cycles, at points $t=m/(n+1)$ with $m=1..n$ ). Igor measures a complete set of stabiliser measurements and applies error correction based on the error syndrome. We take Igor's action to be instantaneous although it is of course trivial to assign it a finite time $\delta$ as indicated in Table~\ref{table:first_criterion}. If Igor's analysis indicates that a correction is necessary, then the appropriate correction is applied {\it perfectly} -- this is a proxy for the reality that one can simply note the need for correction and update future operations to allow for it, thus never needing to apply an imperfect physical operation to the identified qubit. Note however that altering this principle to instead apply a noisy fix would make negligible difference to the observed integrity, since it is merely one additional operation for Igor's circuit which at minimum involves over a dozen gates.

Presently we evaluate the integrity metric for three different well-known codes: the five-qubit code which is the smallest possible error correcting code~\cite{laflamme1996perfect}, the seven-qubit Steane code~\cite{divincenzo1996fault} and the nine-qubit surface code~\cite{dennis2002topological}. We will compare the inherent properties of these codes, both in their simple and fault-tolerant variants, and we will show examples where the various milestones described in the previous section are (or are not) met. 

\begin{figure}
\centering
\includegraphics[width=1\linewidth]{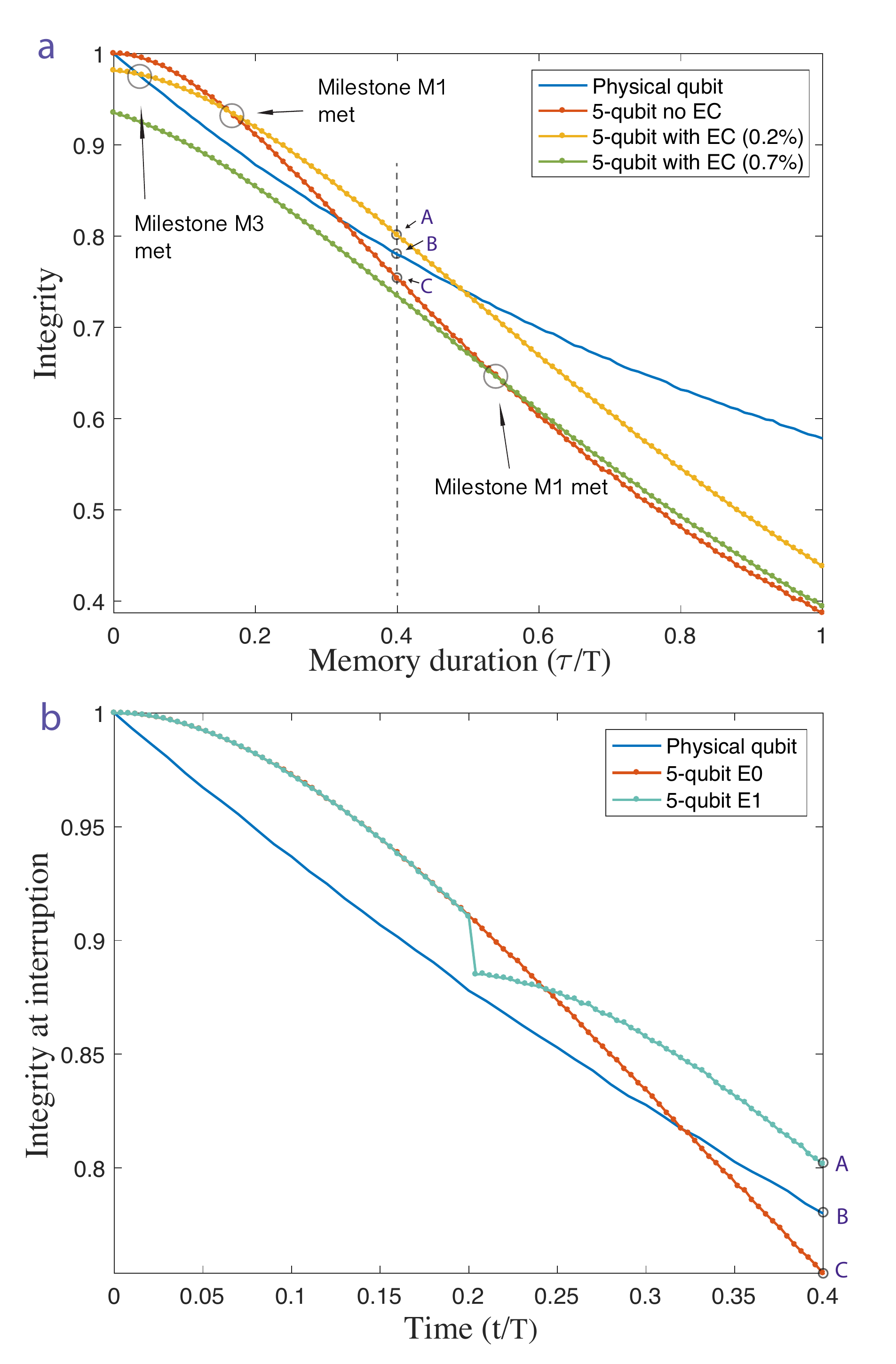}
\caption{{\bf The integrity of different types of memory channel (upper), and the integrity change during a given memory channel (lower). } (a) Memory integrity assessed over many different durations $\tau$ ranging from zero to $T$, the single-qubit decoherence time. Blue line: A single physical qubit, i.e. no encoding. Red: a memory using the five-qubit code for the stored qubit, but without active correction during the channel. Orange: the same five-qubit code, but now with a round of error correction performed mid-way through the memory duration, i.e. at time $t=\tau/2$. Error rate in operations during the correction cycle is $0.2\%$. Green: As for orange, but error rate $0.7\%$. (b) We plot the `integrity at interruption' in order to look inside three specific memory channels during their operation. The three channels all have duration $\tau=0.4\,T.$ As explained in the text, the interesting feature is the step-like decline occurring at $t=\tau/2=0.2\,T$ when Igor performs imperfect error correction.}
\label{fig:multi-1}
\end{figure}

We begin by explaining the nature of the graphs shown in this section. Typically they are of the general form exemplified by Fig.~\ref{fig:multi-1}(a). On the vertical axis we show the integrity, as defined earlier, which of course is equal to unity for an ideal memory. The horizontal axis shows the duration $\tau$ of the memory channel(s) in question; the duration is shown as a ratio with respect to the environmental decoherence rate $T$ which is the decoherence time of an isolated single physical qubit (see error model specification in Appendix~\ref{appendix:errorModel}). Each point along a curve in the figure is thus the integrity of a specific kind of memory when operating for the specific duration indicated by the horizontal axis. 

There are four types of memory channels shown in Fig.~\ref{fig:multi-1}(a): The simple one-qubit memory $\Theta$ (shown in blue), an encoded channels without Igor's error correction, $\Phi^0$ (red), and two channels where Igor does perform one round of correction $\Phi^1$ (yellow, green). The last two differ only in that Igor's error correction circuits have error rates $0.2\%$ and $0.7\%$, respectively. In all cases the encoded channels are using the five-qubit code. Encoding and decoding tasks, performed by Alice and Bob, are perfect as per the definition of integrity (we defer the discussion of noisy Alice and Bob, an unavoidable reality in real benchmarking experiments, to later in this section). 

Igor's error rate of $0.2\%$ is low enough for him to perform well and consequently we observe two desirable line crossings in the figure. For all durations $\tau > 0.16 \,T$ we see that the error corrected memory $\Phi^1$ (yellow) is superior to the un-corrected memory $\Phi^0$ (red). Thus for any $\tau > 0.16 \,T$ we meet the {\textit{M1: beneficial error correction}} milestone specified earlier in Eqn.~(\ref{eq:IgorIsGood}). Furthermore, for all durations $\tau$ between $\tau\simeq0.035\,T$ and $\tau\simeq0.49\,T$ the corrected memory $\Phi^1$ (yellow) has superior integrity to the single-qubit memory $\Theta$ (blue).  Note however that in comparing the single-qubit channel $\Theta$ to the encoded channels, we have not introduced any scaling factor to adjust their relative durations (i.e. we have set $\alpha=1$ in Eqn.~(\ref{eq:versusSingleQubitMem})). This might be considered unreasonable unless the physical platform embodying the memory system is capable, in principle, of performing transversal gates in one step so that operations on logical qubits are on the same timescale as operations on physical qubits. With this important caveat, we can say that for any duration in the range $0.035\,T<\tau<0.49\,T$ we can meet the {\textit{M3: beneficial encoded memory}} milestone, Eqn.~(\ref{eq:versusSingleQubitMem}).

In order to discover whether we can meet the remaining two milestones, and in particular the highly-desirable \textit{M4: strictly superior encoded memory} milestone, we would need to consider channels with multiple rounds of error correction; this is shown presently. 

The green line, corresponding to the higher per-gate error rate of $0.7\%$ during Igor's error correction, never surpasses the integrity of the single physical qubit memory; therefore this channel does not meet milestone {\textit{M3: beneficial encoded memory}}. However when the duration $\tau>0.55\,T$ it does (barely) surpass the integrity of the $\Phi^0$ channel. Therefore milestone {\textit{M3: beneficial error correction}} is met.

Three points labelled $A$, $B$, and $C$ have been highlighted in Fig.~\ref{fig:multi-1}(a). They lie at a value of the duration, $\tau=0.4\,T$ for which the high-fidelity corrected channel is superior to the single physical qubit memory $\Theta$, which in turn is superior to the encoded-but-uncorrected channel $\Phi^0$. One might wish to understand how the integrity varies {\it over the course} of the duration of those memory channels. In fact, the question is not entirely proper since integrity is only defined as a property of the entire channel; but we can ask what would happen if Bob were to `step in early' at any time between $t=0$ and $t=\tau=0.4\,T$. We suppose that Bob would perform his usual decoding, measurement and guess using the state of the memory system at that premature point. From his performance we can infer an `integrity so far', so to speak, which we might also call the `integrity at interruption'.  Fig.~\ref{fig:multi-1}(b) shows this quantity. The blue and red lines, which correspond to the single-qubit memory $\Theta$ and the encoded memory without correction $\Phi^0$, do not reveal anything interesting. Indeed they precisely correspond to the same lines in the region $0<\tau<0.4\,T$ in the upper panel (in effect, we have just `zoomed in'). For these two cases the noise on the memory is simply a continuous process; when Bob interrupts our channel that should have had duration $\tau=0.4\,T$, it is exactly equivalent to having a memory of the shorter duration. 

The green line in Fig.~\ref{fig:multi-1}(b) corresponding to the error-corrected channel  $\Phi^1$ is far more interesting. It is exactly coincident with the $\Phi^0$ line until $t=0.2\,T$ because in these cases Bob interrupts before Igor performs his error correction cycle. But then the `integrity at interruption' falls sharply, i.e. there is a significant difference between Bob interrupting immediately before Igor's effort, versus doing so immediately afterwards. The reason is that Bob's process of decoding the memory begins with a round of error correction and {\it his} error correction is perfect; thus it can only be worse to have Igor apply his own flawed effort at correction immediately beforehand. However despite the sharp step down, the eventual integrity at full duration is higher. This is because Igor's efforts have reset the accumulation of errors, lowering the overall chance of an uncorrectable set of errors (i.e. 2 or more errors, for the five-qubit code) over the course of the complete memory channel. We see this evidenced by the inverted parabolic curve immediately after Igor's action: in effect the environment must `start again' to build up significant probability of weight-2 errors.

It is interesting to reflect further on the observation that error correction cannot increase the quantity `integrity at interruption', assuming we have no knowledge of the initial encoded qubit (given such knowledge we can trivially increase integrity by erasing the memory and reinitialising it). Any form of error correction, with whatever code and however well performed, is a process that merely `delays the inevitable' in the sense that integrity must fall; we can only alter the rate at which it falls. For the ultimate goal of fault tolerant quantum computing, we must slow the decay of integrity to such an extent that the entire calculation can take place before an error becomes likely. The fact that integrity is a non-increasing function of time is a merit versus over other measures (such as the simple fidelity with respect to an ideal state) which can both fall and rise, so creating the false impression that quantum information is somehow being  regenerated.

\begin{figure}
\centering
\includegraphics[width=1\linewidth]{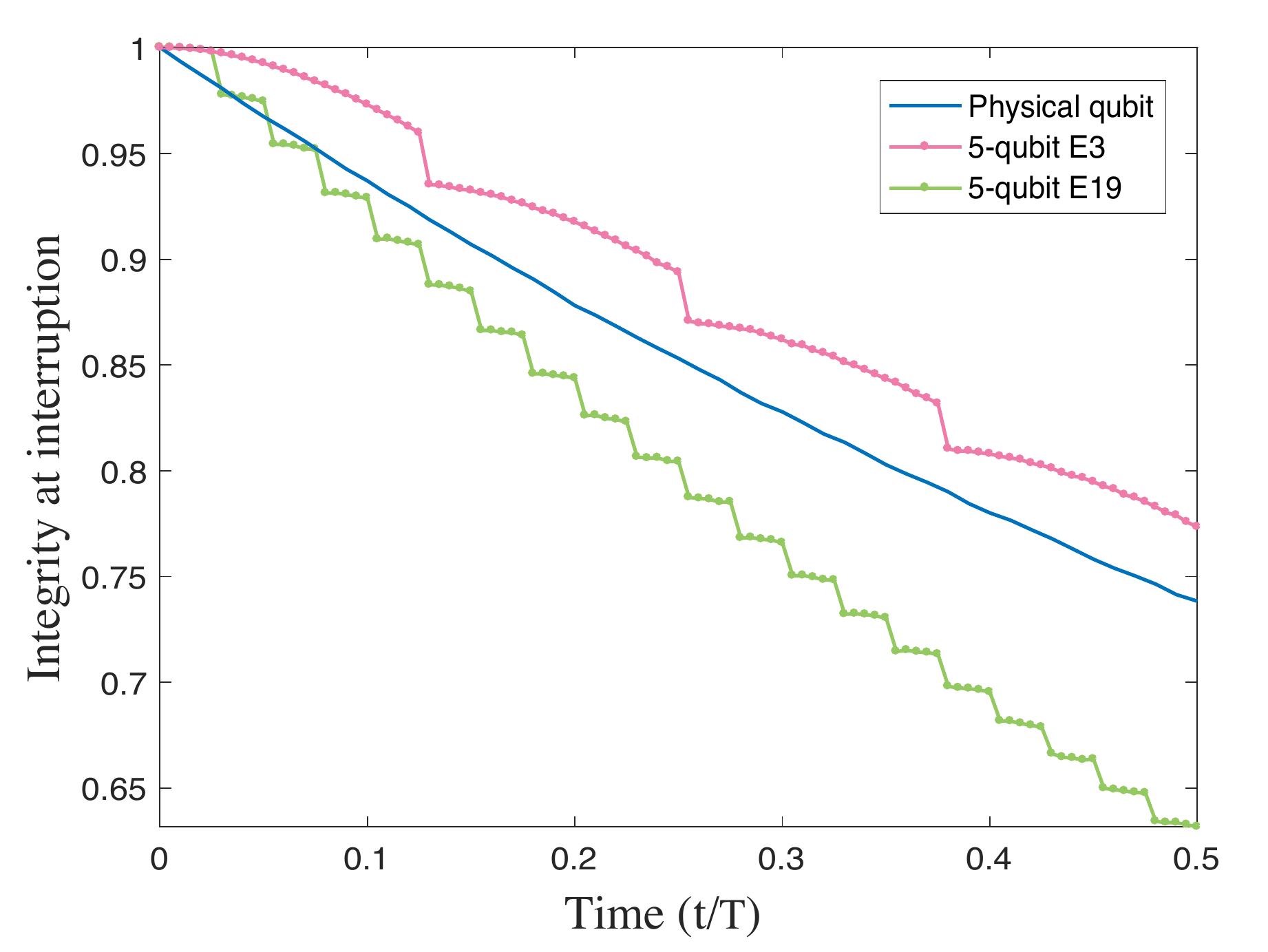}
\caption{{\bf Integrity change during three different memories, each of duration $\tau=0.5T$.} This figure is equivalent to Fig.~\ref{fig:multi-1}(b), but with three rounds (pink) or nineteen rounds (green) of error correction (gates error rate 0.2\%) applied during the period of memory storage. Clearly three rounds of error correction sustains the logical qubit while too many rounds corrupt the logical qubit. }
\label{fig:multi}
\end{figure}

A related observation is the following: The rate at which we should apply error correction cycles has some optimum which depends on the relative severity of environmental decoherence per unit time versus the error rate within our error correction process (the noise in Igor's circuits). We should not apply error correction more frequently than this rate, or else the loss of the integrity will be dominated by the noise we introduce in our error correction cycles. This is made apparent by the simulation results shown in Fig.~(\ref{fig:multi}) where we again plot the `integrity at interruption' as in Fig.~\ref{fig:multi-1}(b), but now for three different channels of common duration $\tau=0.5\,T$, the channels being the single qubit memory $\Theta$, and memories using three or nineteen error correction cycles ($\Phi^3$ and $\Phi^{19}$). From the right hand side of the graph we find the integrities  of the three memory channels: they are approximately $0.74$, $0.78$ and $0.63$ respectively, i.e. the memory channel featuring nineteen correction cycles is by far the worst, while three cycles (which is in fact the optimum here) provide a superior integrity versus the single qubit memory. The reason is clear from inspecting the curves: the `integrity at interruption' reveals that the decay of the over-corrected channel is indeed dominated by the step-like drops associated with noise from Igor. 

With that introduction, we now present a series of simulations which contrast different codes, and also compare fault-tolerant versus non-fault-tolerant implementations of error correction circuits. Unless otherwise noted we use the standard error model of homogeneous  Pauli noise occurring without correlation, and for Igor's circuits the noise occurs on all circuit operations with equal probability. It is worth stressing that the relative performance of the codes may differ greatly when this error model is substantially varied.

\begin{figure}
\centering
\includegraphics[width=1.05\linewidth]{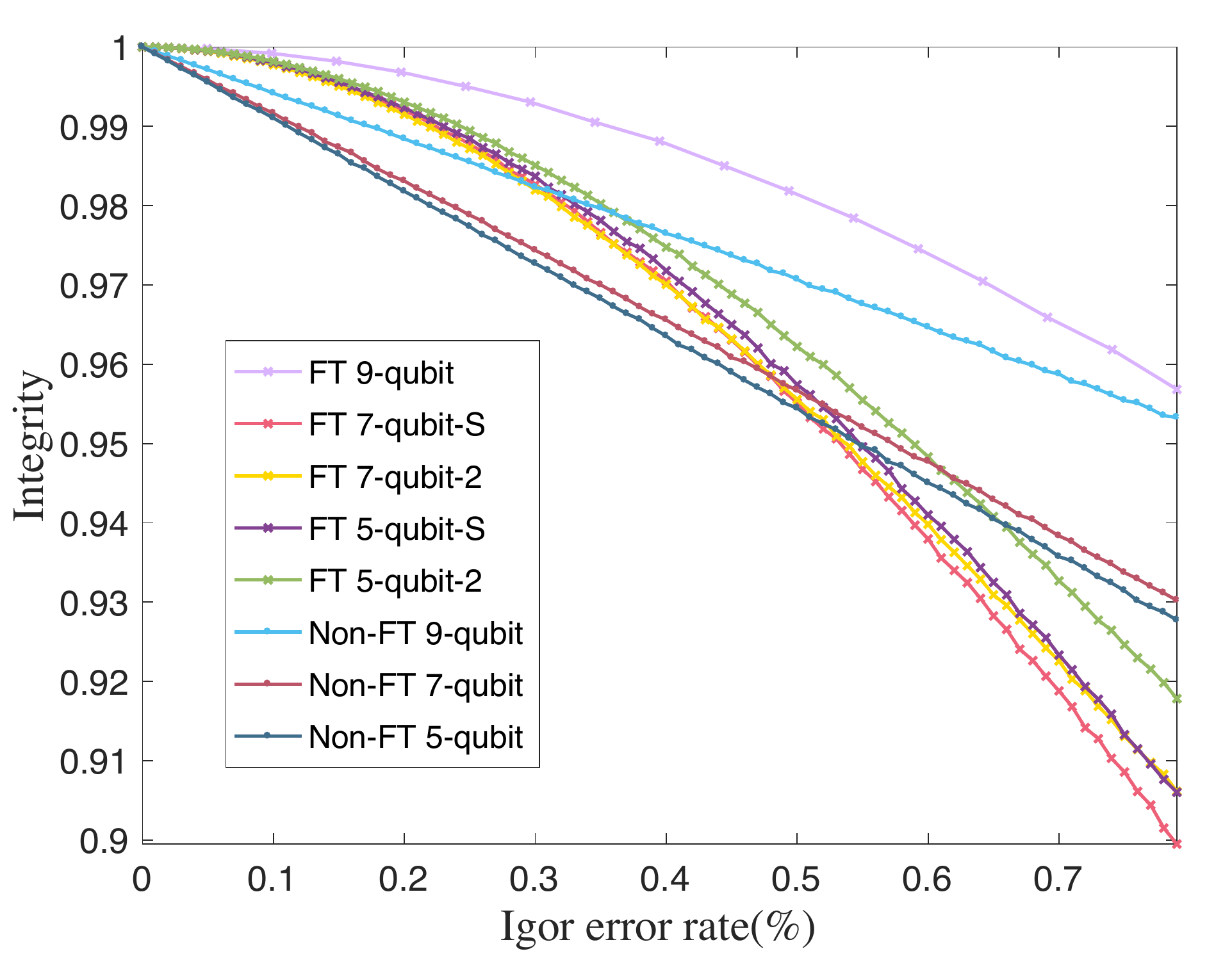}
\caption{{\bf Integrity change with increasing gate error rate. } Here the duration of our memory is set to zero, in order to directly inspect the negative impact of an imperfect error correction performed by Igor. The horizontal axis shows the level of noise associated with each circuit element of Igor's circuits. We analyse memories based on the five-qubit, the Steane, and the nine-qubit codes. Igor's error correction is performed either in a simple, non-fault tolerant fashion or with full fault tolerance. As explained in the text, the various line shapes and the relative levels of performance are straightforward to understand qualitatively. }
\label{fig:IgorError}
\end{figure}

The appendix shows the various encoding, decoding, and error correction circuits which we use in the simulations described here. As a first step toward comparing the efficacy of different codes, we begin by reporting a special case which is achieved by setting the memory duration to zero, and simply investigating the impact of the error correction process itself. Thus, we take a perfectly encoded qubit prepared by Alice and present it directly to Igor who performs an (entirely unnecessary!) error correction cycle before passing the encoded qubit directly to Bob for his analysis. The reduction in integrity is thus purely due to Igor's action.  The results are shown in Fig.~\ref{fig:IgorError}. Notice that in contrast to all other figures in this paper, the horizontal axis here is not time (since the duration is zero) but rather Igor's error rate. 

Figure~\ref{fig:IgorError} includes eight different options for the encoding and correction of a logical qubit. Three different codes are considered: the five-qubit code, the seven-qubit Steane code (which is also the smallest 2D color code), and the nine-qubit surface code. For each of these, the performance of a non-fault-tolerant (non-FT) Igor is plotted. For the nine-qubit code, a second curve shows the performance when Igor employs a specific FT error correction circuit (see Fig.~\ref{fig:FTScheme}(e)). For each of the other two codes, we display the performance of two different FT circuits: The standard `Shor' approach using four ancilla qubits, and an alternative method very recently proposed by Chao and Reichardt~\cite{chao2017quantum} which requires only two ancillas, see Fig.~\ref{fig:FTScheme} panels (c) and (d).

There are several interesting general observations to be made from Fig.~\ref{fig:IgorError}. Firstly, it is reassuring to note that two logically-necessary features are indeed present: One observes that all the cases which employ non-fault-tolerant (non-FT) error correction for Igor have the expected linear decay as Igor's error probability $p$ increases from zero: integrity goes as $1-c\,p$ for some constant $c$ because non-FT circuits are vulnerable to single errors. Meanwhile the scenarios featuring FT error correction all have the expected inverted-parabolic shape: the integrity goes approximately as $1-k\,p^2$ when Igor's error probability $p$ is small. Circuits of this kind are `immune' to single errors and vulnerable only to weight two (or higher) errors. Note that for higher (but still sub-$1\%$) error rates for Igor, the fault tolerant circuits become inferior to the simpler non-FT circuits. The reason is essentially combinatorial scaling: the FT-circuits are generally considerably more complex with far more gates, thus as gate failure probability $p$ increases the risk of a double error in these complex circuits eventually outweighs the risk of a single error in the simple non-FT circuits. Thus one should not suppose that `fault tolerant circuits are always better' -- for small codes and appreciable rates of gate error, they may not be.

The different gradients in the various linear and parabolic curves can be qualitatively understood by considering two desiderata. The first is the {portion} of all possible weight-2 errors that actually prove to be correctable. For example, the five-qubit code is corrupted by all weight-2 errors, but the seven-qubit Steane code can correct any pair of errors if (and only if) one is of type $X$ and one of type $Z$. The nine-qubit surface code has the highest portion of `harmless' weight-2 errors in this sense. The relative ordering of the non-FT codes can be explained by this feature alone. However for the FT codes, there is another competing feature: as noted above the complexity of the FT error correction circuits is what `kills' their performance, so simpler circuits are superior. Consistent with this principle, we see wherever an appreciable performance gap exists between the `two-ancilla' variant of a FT code versus the `Shor' variant of the same code, the former is always superior. Moreover the FT circuits for the five-qubit code are more simple than those the seven-qubit Steane, thus among the FT curves the five-qubit outperforms the Steane. 
Remarkably FT circuits for the nine-qubit code exist which are actually very simple (as previous authors have noted~\cite{svore2014,wootton2016noise}),  and thus the FT surface code benefits from both desirable features described here, and is unconditionally superior to all other codes in the plotted error range. However, it does require the largest number of qubits: the 9 data-qubits themselves, and Igor also requires 6 ancillas in order to perform stabiliser evaluation without error propagation. Thus one might argue that the FT five-qubit code, in its two-ancilla variant, provides better `value per qubit' since it requires a device with only 7 qubits in total.

\begin{figure}
\centering
\includegraphics[width=0.9\linewidth]{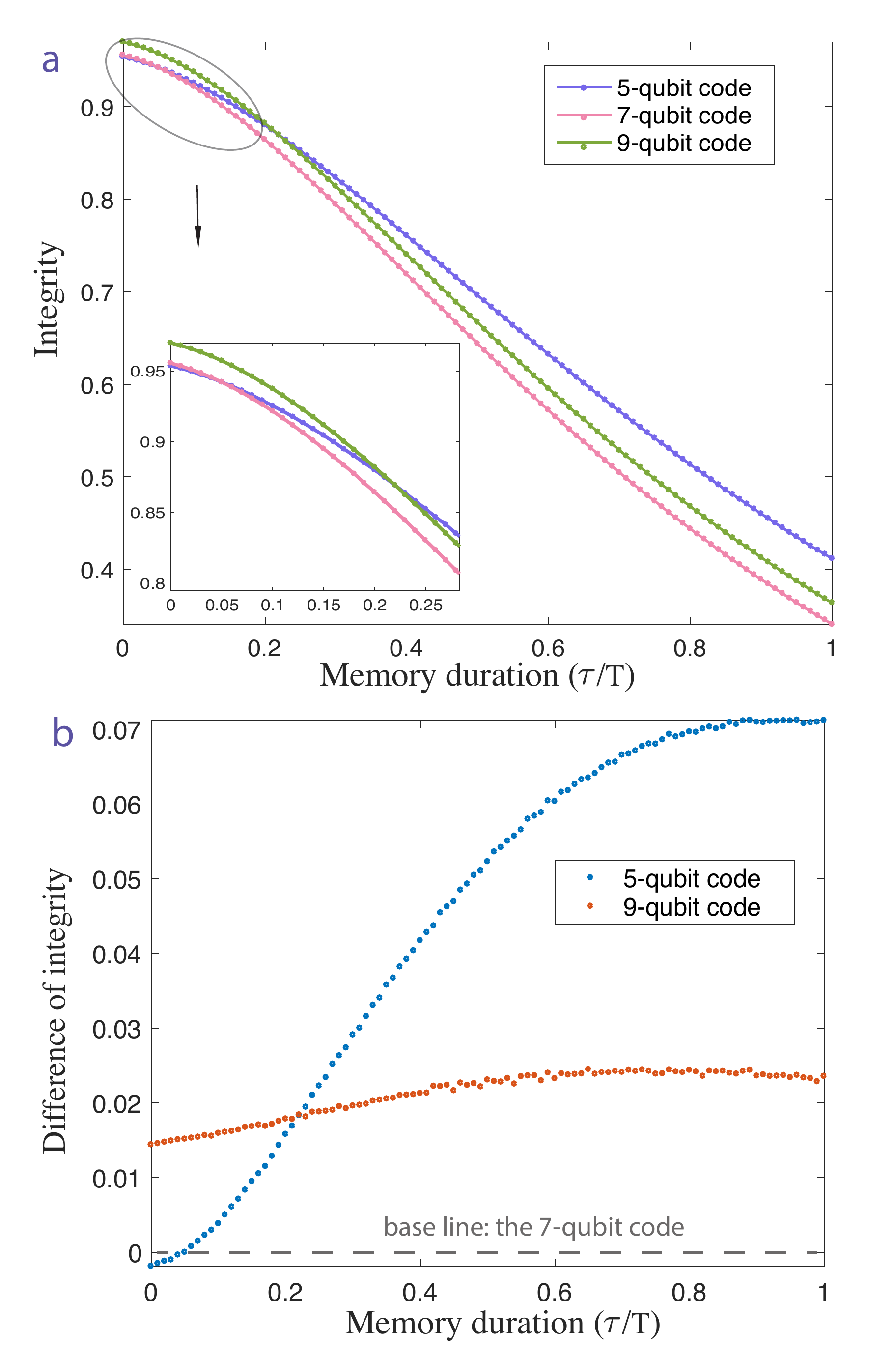}
\caption{{\bf Comparison between the 5/7/9 qubit codes.} The data shown are for our canonical Alice-Igor-Bob scenario where total memory duration $\tau$ during which pure environmental decoherence occurs continuously and a single (imperfect) round of error correction occurs midway at $t=\tau/2$. See e.g. Fig.~\ref{fig:5qubitScheme}. The error rates used for all the gate operations during the error correction procedure are 0.5\%. The lower panel (b) presents the same data but now with respect to the the Steane code performance, so that the integrity of that channel now lies along the horizontal axis.}
\label{fig:comparionEncoded}
\end{figure}

\begin{figure}
\centering
\includegraphics[width=0.9\linewidth]{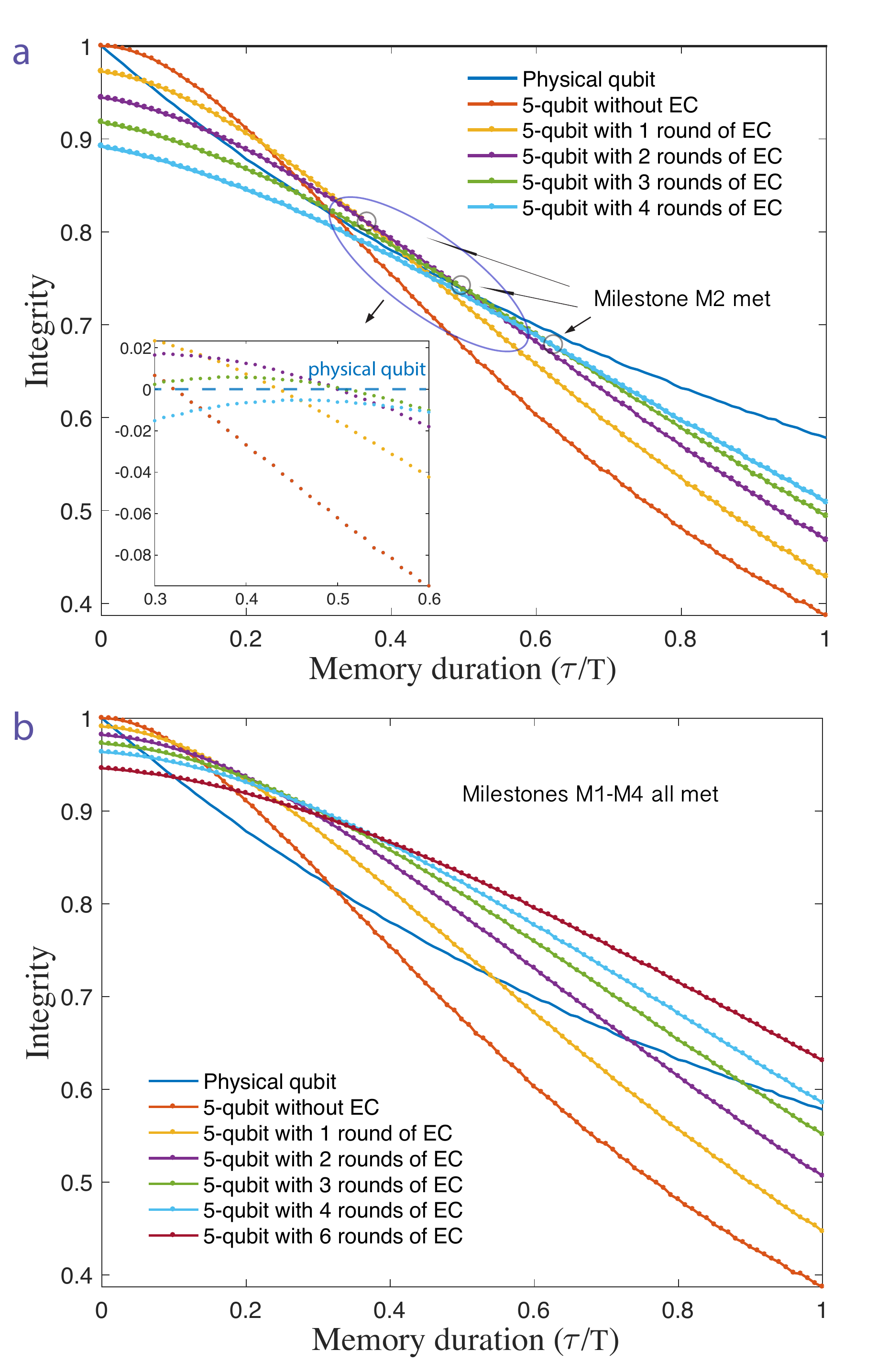}
\caption{{\bf Multiple rounds of quantum error correction(EC).} The integrity of a family of memory channels all employing the five-qubit code but  differing in the number $n$ of rounds of error-correction performed during the memory, where $n=0,1,2,3,4$ or $6$. Our imperfect agent Igor performs error correction cycles at times $t=m\tau/(n+1)$ for $m=1..n$. In the upper panel (a) Igor's gate-level error rate is $0.3\%$. As explained in the main text, the system meets milestones M1, M2 and M3 but fails to meet M4. In the lower panel (b) Igor's error rate is now $0.1\%$ and we see that by choosing a suitable $n$ we can select a five-qubit encoded memory that will beat the single-qubit memory for any desired duration $\tau$, so meeting milestone M4: Strictly superior encoded memory.}
\label{fig:multiigor}
\end{figure}

It is important to remember that the comparison made in Fig.~\ref{fig:IgorError} is for zero environmental error. The relative performance of different codes will change once we deploy them properly into a memory channel where environmental noise is degrading the encoded qubit. Figure~\ref{fig:comparionEncoded} shows the integrity change of the memory under our standard memory channel scenario $\Phi^1$ i.e. `one use of Igor's error correction midway' where the code employed is either the five-qubit, Steane, or nine-qubit code (all with non-FT correction). See Fig.~\ref{fig:5qubitScheme} for the explicit circuit used in the five-qubit code case; circuits for the other cases differ simply by substituting the appropriate stabiliser checks. All curves in this figure correspond to an internal error rate for Igor's operations of $0.5\%$. Thus the far left of the figure, with $\tau=0$, gives us the same set of three data points as can be read from Fig.~\ref{fig:IgorError} when the $x-$axis, the error rate, is $0.5\%$. We see that the Steane code is marginally superior to the five-qubit code, but both are markedly inferior to the nine-qubit code. However, as we move away from the hard left of Fig.~\ref{fig:comparionEncoded} to consider increasing duration of the memory, we find that the five-qubit code surpasses first the Steane and then even the nine-qubit code. The reason is that as more environmental error accrues, a code with a larger number of physical qubits will reach the point where two-or-more errors are present, i.e. the situation where the logical qubit may be corruptted, at an earlier time. 

In preceding figures we have focused on cases where a single round of error correction is applied during a memory channel, and we have identified points where our milestones M1 and M3, would be satisfied. In Figure~\ref{fig:multiigor} we show how the use of multiple rounds of error correction (equispaced within the duration of the memory channel) may allow us to meet milestone {\it M2: Beneficial multi-round error correction} associated with  Eqn.~(\ref{eq:nRounds}), or even milestone {\it {M4: Strictly superior encoded memory}} associated with Eqn.~(\ref{eq:unconditionallyBetter}). In the upper panel, Igor's error rate suffices for the former but not the latter; in the lower panel Igor's error rate is set to $0.1\%$ which proves to be sufficient to achieve the fourth milestone.

Before concluding this comparison of different codes, we should stress that our intention is not to identify ``best and worst'' codes but rather to show the circumstances in which various codes can be the better choice. We also recognise that there are other merits beyond the question of how well a code preserves channel integrity -- for example, the Steane code (which is also the smallest instance of the 2D color code) has the significant merit versus the smaller five-qubit that all Clifford operations can be applied transversally.

\section{Assessing integrity in a real experiment}\label{sec:realExp}

In all the theory and the numerical simulations described above, Alice and Bob are perfect agents: they provide the framework within which we assess the memory channel. However, if we are to assess integrity in an experiment -- i.e. if it is to be a practical measure for benchmarking quantum memories -- then we must tackle the reality that Alice and Bob are merely phases of an experiment within which all operations are imperfect. Bearing this in mind, to what extent can one can still assign an integrity to the memory channel? And more importantly can we still confidently assert that the integrity of one channel is superior to another, in order to determine whether milestones such as those identified in Section~\ref{milestones} have been met?

In the simulations reported in this section, we apply the same error model and error severity to the actions of Alice and Bob, as we do to Igor's error correction cycle. It is crucial now to specify the particular circuits that Alice and Bob use to perform their functions (whereas before, since they were perfect agents, any circuit performing the desired function was equivalent). 

In the idealised case we spoke of Alice preparing any encoded qubit she wished, i.e. she  used a general encoding circuit such as those displayed in Fig.~\ref{fig:circuitsRandom}. Bob used a complex procedure involving a full round of error correction followed by inverting Alice's general encoder to map an arbitrary encoded qubit back to a single physical qubit. However, the definition of integrity corresponds to Bob's performance when Alice opts for the worst possible choice of qubit state to encode (or rather, when she picks between the two states $\ket{\psi}$ and $\ket{\psi_\perp}$, which Bob has the most difficulty differentiating post-memory). If we have foreknowledge of which states these are, we need only find circuits for Alice and Bob to use which perform {\it equivalently} to their general purpose circuits in these special cases. Fortunately for a broad family of error models (see Appendix~\ref{appendix:whenIsPauliEnough}) we know that the worst case choice Alice can make will correspond to Pauli basis states, i.e. $\{\ket{\psi},\ket{\psi_\perp}\}$ will be either $\{\ket{0},\ket{1}\}$ or $\{\ket{+},\ket{-}\}$ or $\{\ket{y+},\ket{y-}\}$. Our challenge is therefore to find specific encoder circuits for Alice and analysis circuits for Bob for these special cases. This must be done in such a way that we recover the ideal performance of Alice and Bob when they are indeed error-free, but we obtain best-possible performance for Alice and Bob when they are error-burdened. In short, we look for compact fault-tolerant realisations of Alice and Bob for the cases where $\ket{\psi}$ and $\ket{\psi_\perp}$ are Pauli basis states. 

We emphasise that once we equip Alice and Bob with suitable circuits, we have a full prescription for an experimental test of integrity: the experimental protocol is simply to follow the process listed in Tables~\ref{table:basic_protocol} and \ref{table:first_criterion}, with the sole modification that Alice randomly picks between Pauli eigenstates, and given this pick both her encoding circuit and Bob's decoder are selected accordingly from optimised circuits such as those in Fig.~\ref{fig:circuitsZero}. Thus integrity is evaluated without state tomography. 

\begin{figure}
\centering
\includegraphics[width=0.9\linewidth]{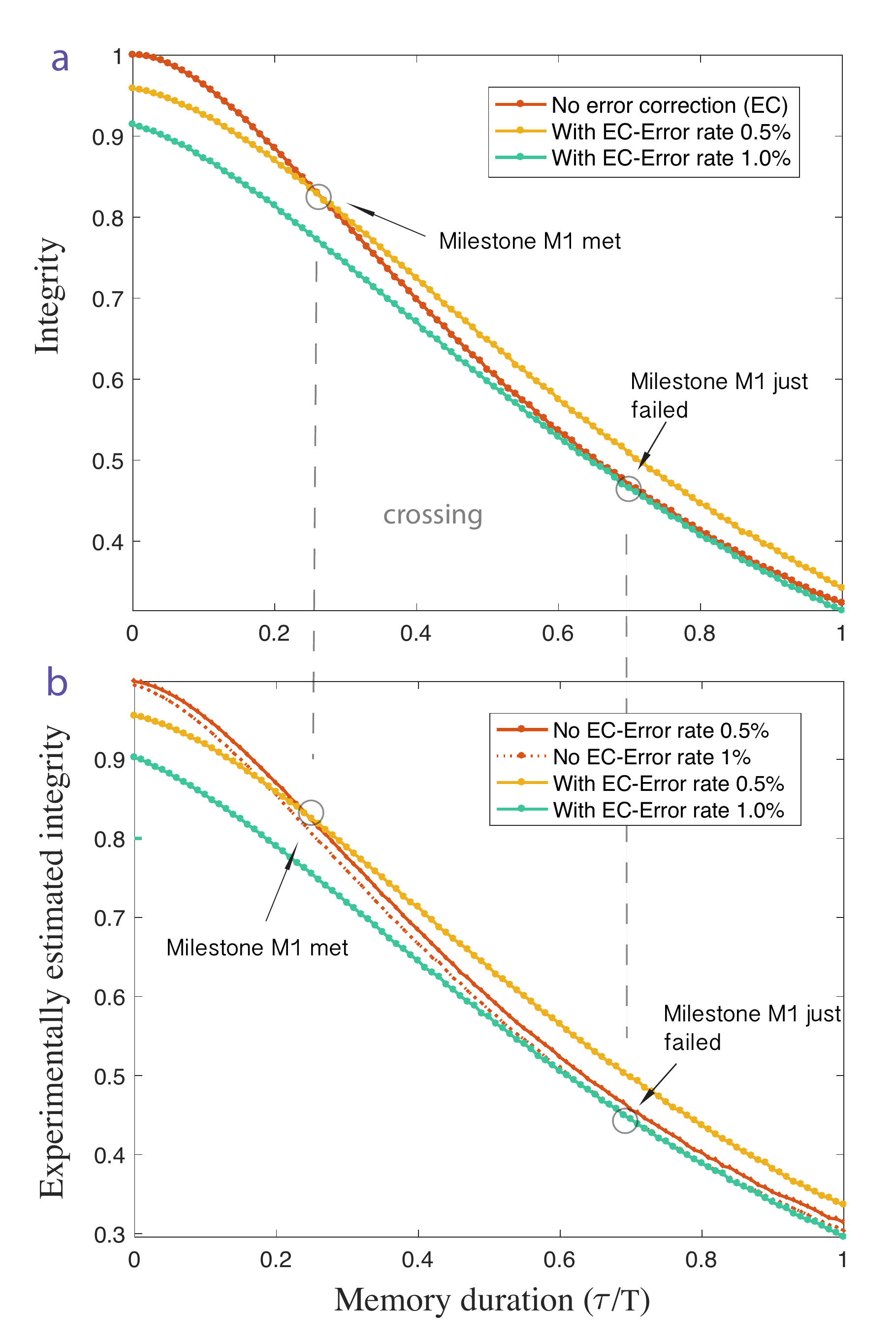}
\caption{{\bf Imperfect, but fault tolerant, Alice and Bob.} The scenario in the upper panel (a) corresponds to three different memory channels each using the Steane code to protect information. The encoder Alice and the analyser Bob are both ideal, as required in the definition of integrity. We mark the meaningful line crossing which corresponds to meeting milestone M1 (for the orange line) or just failing to do so (blue line). In the lower panel (b) we present the same analysis but now with errors during Alice and Bob's circuits at the same level as Igor's. Specifically, Alice uses the circuits shown in Fig.~\ref{fig:circuitsZero} to encode her qubits into $|0\rangle_L$. Fig.~\ref{fig:circuitsZero} also shows how Bob differentiates between $|0\rangle_L$ and $|1\rangle_L$ by simply measuring all qubits in the $z$-basis, performing classical error correction, and checking the parity of a certain subset.
Note that the key crossing (and failure to cross) from the upper panel are well approximated in the lower, indicating that experimental evaluation of integrity is achievable.}
\label{fig:sevennoisyAB}
\end{figure}

\begin{figure}
\centering
\includegraphics[width=0.9\linewidth]{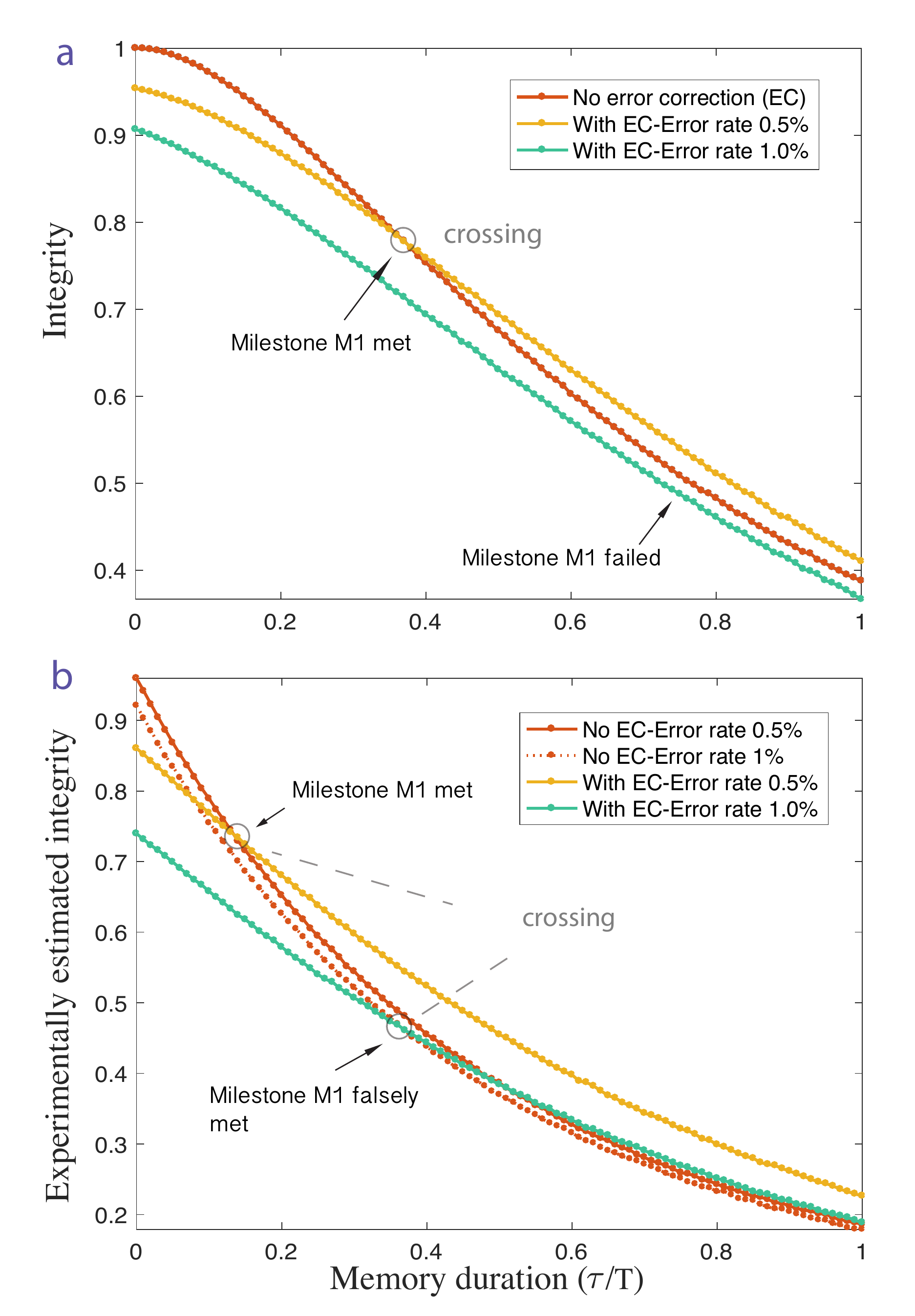}
\caption{{\bf Imperfect, and non fault tolerant, Alice and Bob.} The comparison made here is similar to that in Fig.~\ref{fig:sevennoisyAB} except that now the memory channels employ the five-qubit code and moreover in the lower panel (b) the circuits used by Alice and Bob are not fault tolerant. Here, Alice uses the circuits specified in Fig.~\ref{fig:circuitsZero} to encode qubits into $|-\rangle_L$ and Bob uses the circuit shown in that figure to differentiate between $|+\rangle_L$ and $|-\rangle_L$.
In contrast to Fig.~\ref{fig:sevennoisyAB} there is now a profound difference between two panels and one could not directly assert that a line crossing in the lower panel implies a crossing would exist in the upper panel.}
\label{fig:fivenoisyAB}
\end{figure}

The data plotted in Fig.~\ref{fig:sevennoisyAB} and Fig.~\ref{fig:fivenoisyAB} show the effect of allowing Alice and Bob to become noisy, for the Steane code and the five-qubit code respectively. Circuit details are given in the captions.

For Steane code(Fig.~\ref{fig:sevennoisyAB}) we observe an excellent agreement between the `true' integrity that would be measured if one were able to use ideal agents Alice and Bob, and the estimate of the integrity that results from using imperfect agents. We note that there is only a slight variation in the location of the crossing point, and that if a crossing occurs in the true integrity (as for the case when Igor's error is $0.5\%$) then a crossing also occurs in the estimate; the specific crossing shown here is that which would show milestone M1 has been met. Conversely when a crossing does not occur in the true integrity, it also fails in the estimate (here, for the case when Igor's error is $1\%$). The reason for the excellent agreement is that both Alice and Bob's circuits are robust against errors. Note we adopt the recently proposed protocol from Ref.~\cite{goto2016minimizing} -- by using a single additional qubit in her encoding process, Alice is able to detect many errors, and if such errors are detected the  encoding process is restarted again until no error is detected. This we are free to do since Alice is `not on trial here' so to speak; our goal is to fairly evaluate the memory channel involving the environmental noise and Igor's imperfect attempt(s) at error correction. Similarly, it would be legitimate to employ circuits for Bob which reject some outcomes completely and do not count them towards the estimate of his guess success rate, if those cases definitely correspond to some failure within Bob's own processes. An example would be, if a measuring device fails to return any result at all.

In our second example of noisy Alice and Bob, shown in Fig.~\ref{fig:fivenoisyAB} we employ the five-qubit code and, crucially, we do not employ fault tolerant procedures for Alice and Bob. For the five-qubit code there is relatively little literature describing fault tolerant state preparation and measurement (in contrast to the Steane code where there are numerous circuits exhibited in the literature, and progress~\cite{goto2016minimizing} has been made as recently as 2016). Moreover the smaller size of the five-qubit code itself may mean that it is targeted by the very earliest experiments where the additional complexity associated with making Alice and Bob fault tolerant is an unwelcome obstacle. Unfortunately, when the tasks performed by Alice and Bob become vulnerable to single gate failures, the resulting memory integrity estimates become very poor approximations to the true integrity.  In the lower panel of Fig.~\ref{fig:fivenoisyAB} we see that the line shapes have changed, losing the inverse-parabola shape for short memory durations. We do still see line crossings, but they occur at significantly different locations. Most troublingly, a line crossing can occur in the experimental data when no such crossing {\it would occur} if Alice and Bob were ideal. Thus, the observation of a crossing in the data is not, in of itself, strong evidence that the actual memory channel has met a meaningful milestone (such as {\it M1: Beneficial error correction} in this case).

Despite these issues, it can be possible to make use of data such as that in Fig.~\ref{fig:fivenoisyAB}. One would need to perform additional theoretical analysis in order to justify the claim that any observed crossing is indeed meaningful. For example, if the errors in the various circuit elements are well characterised then one could perform simulations equivalent to those presented in this paper. Essentially one would produce a version of Fig.~\ref{fig:fivenoisyAB}(b) calculated with an accurate error model in order to compare with the observed data; if the match proved to be good, one could use further simulation to discover the integrity that {\it would have been observed} with ideal Alice and Bob. In order words, if the data closely matches a simulation such as  Fig.~\ref{fig:fivenoisyAB}(b), one might fairly state that this is strong evidence that the integrity is as shown in Fig.~\ref{fig:fivenoisyAB}(a). 

In summary, we can say that integrity can be assessed experimentally in a straightforward protocol: Acting as Alice we choose a qubit state then we perform a series of experimental runs where each run ends in a measurement from which, as Bob, we `guess' the original state with the binary outcome `succeeded' or `failed'. We continue until we have a good estimate of Bob's probability of success $p_g$; if the system is such that $p_g$ depends on Alice's choice, then we find the least-favourable choice. The integrity of the memory is then simply $\mathcal R(\Phi)=2p_g-1$. In this section we have shown that the creation of the logical qubit, i.e. Alice's circuit, as well as Bob's analysis circuit, can both be noisy and yet we can obtain an excellent estimate of the integrity of the memory channel itself (factoring out Alice and Bob).

\section{Generalisations}
The analysis presented here has defined the integrity of a memory channel, where that channel stores a single logical qubit. The specific codes we have considered are distance three (a single physical qubit error is correctable) but the definition applies equally to higher distance codes. For cases where a memory channel stores several logical qubits, it is straightforward to generalise our integrity metric: a natural choice for an $m$ logical qubit memory would be to have Alice choose a state of $m$ qubits, encode and transmit to Bob as in our canonical picture (including optionally error correction from Igor) and then Bob decodes and is finally informed of two options -- Alice's true state and a randomly chosen orthogonal state -- between which he must guess. The memory channel's integrity will relate to Bob's worst case performance within this framework.

One can also generalise the notion of integrity beyond memory systems to actual computations. For a single logical qubit the natural generalisation would be to perform multiple transversal gates between the Alice and Bob stages, i.e. in lieu of the pure environmental noise periods. As with the memory channel, this computational process could include one, or more rounds of error correction from our agent Igor.

\section{Conclusion}
To conclude: we have described and assessed a measure called {\it integrity} as a means to benchmark the performance of a code-based quantum memories. Integrity measures how well a memory preserves the distinctiveness of different states. It was introduced recently to assess ion trap based memories in Ref.~~\cite{17BermudezFeasibility}, but is generically applicable to any technology platform. Integrity is a property of the memory channel itself (including any active memory correction routines) independently of the inevitable encoding and measurement stages. Importantly the integrity of a memory can be assessed experimentally in a straightforward manner without the need for full state tomography.  We have identified links between integrity and quantities such as `fidelity of the logical qubit' or the `pseudo-threshold'. 

\section{Acknowledgements}
The authors gratefully acknowledge many useful conversations with our co-authors of the Ref.~\cite{17BermudezFeasibility}, the paper where the Alice-Igor-Bob paradigm was first introduced. These discussions allowed us to set the scope and goals of the present paper. Special thanks go to Markus M\"uller and Alejandro Bermudez for the insightful discussions about about the theoretical merits and characteristics of the integrity measure. 

SCB and XX are supported by the Office of the Director of National Intelligence (ODNI), Intelligence Advanced Research Projects Activity (IARPA), via the U.S. Army Research Office Grant No. W911NF-16-1-0070. The views and conclusions contained herein are those of the authors and should not be interpreted as necessarily representing the official policies or endorsements, either expressed or implied, of the ODNI, IARPA, or the U.S. Government. The U.S. Government is authorized to reproduce and distribute reprints for Governmental purposes notwithstanding any copyright annotation thereon. Any opinions, findings, and conclusions or recommendations expressed in this material are those of the author(s) and do not necessarily reflect the view of the U.S. Army Research Office. 

SCB and NdB are supported by the EPSRC National Hub in Networked Quantum Information Technologies (NQIT.org). The authors would like to acknowledge the use of the University of Oxford Advanced Research Computing (ARC) facility in carrying out this work. http://dx.doi.org/10.5281/zenodo.22558.

%

\appendix

\section{Error model}
\label{appendix:errorModel}

Here we specify the error model used in the numerical simulations presented in this paper. 

Environmental decoherence is modelled as a depolarising process that occurs independently for each physical qubit. Specifically, when our memory qubits are exposed to the environment for some time $t$ then the probabilities of an error is given by 
\[
p=\tfrac{1}{2}\left(1-\exp(t/T)\right).
\]
Given that an error occurs, it is assigned as one of the three Pauli operators $\sigma_X$, $\sigma_Y$, $\sigma_Z$ selected uniformly at random.
 This occurs independently and in parallel for each physical qubit.

Noise also occurs when gate operations are applied by `Igor' while performing error correction cycle(s) during the memory channel in order to actively protect the stored information.  Recall that Alice and Bob, whose actions at $t=0$ and $t=\tau$ frame the memory channel, are considered ideal for the purpose of the definition of integrity; however in the Figs.~\ref{fig:sevennoisyAB} and~\ref{fig:fivenoisyAB}, and the associated main text, we consider the effect of making Alice and Bob as noisy as Igor since this is the likely experimental reality. In all these cases our error model for circuit operations is as follows:
\begin{itemize}
\item A noisy single-qubit gate is modelled by the ideal gate followed, with probability $p_e$, by one of the three Pauli operators $\sigma_X$, $\sigma_Y$, $\sigma_Z$ selected uniformly at random.
\item Noisy state preparation is modelled by ideal preparation followed by a possible error in the same fashion as above.
\item Noisy measurement is modelled by inverting the state to be measured in the relevant measurement basis, with probability $p_e$. So for example, prior to a measurement in the $z$-basis a $\sigma_X$ operation will be applied to the qubit with probability  $p_e$.
\item A noisy two-qubit gate is modelled by the ideal gate followed, with probability $p_e$, with one of the fifteen non-trivial Pauli operators products $I\otimes\sigma_X$, $I\otimes\sigma_Y$,..., $\sigma_Z\otimes\sigma_Z$ selected uniformly at random.
\end{itemize}
Notice that the same error probability $p_e$ is used for all types of circuit element; this is the number that is specified in the main paper as `Igor's error rate' and typically expressed as e.g. $0.3\%$.

\section{Circuits diagrams}
\subsection{Alice and Bob's circuits}

\begin{figure}
\centering
\includegraphics[width=0.8\linewidth]{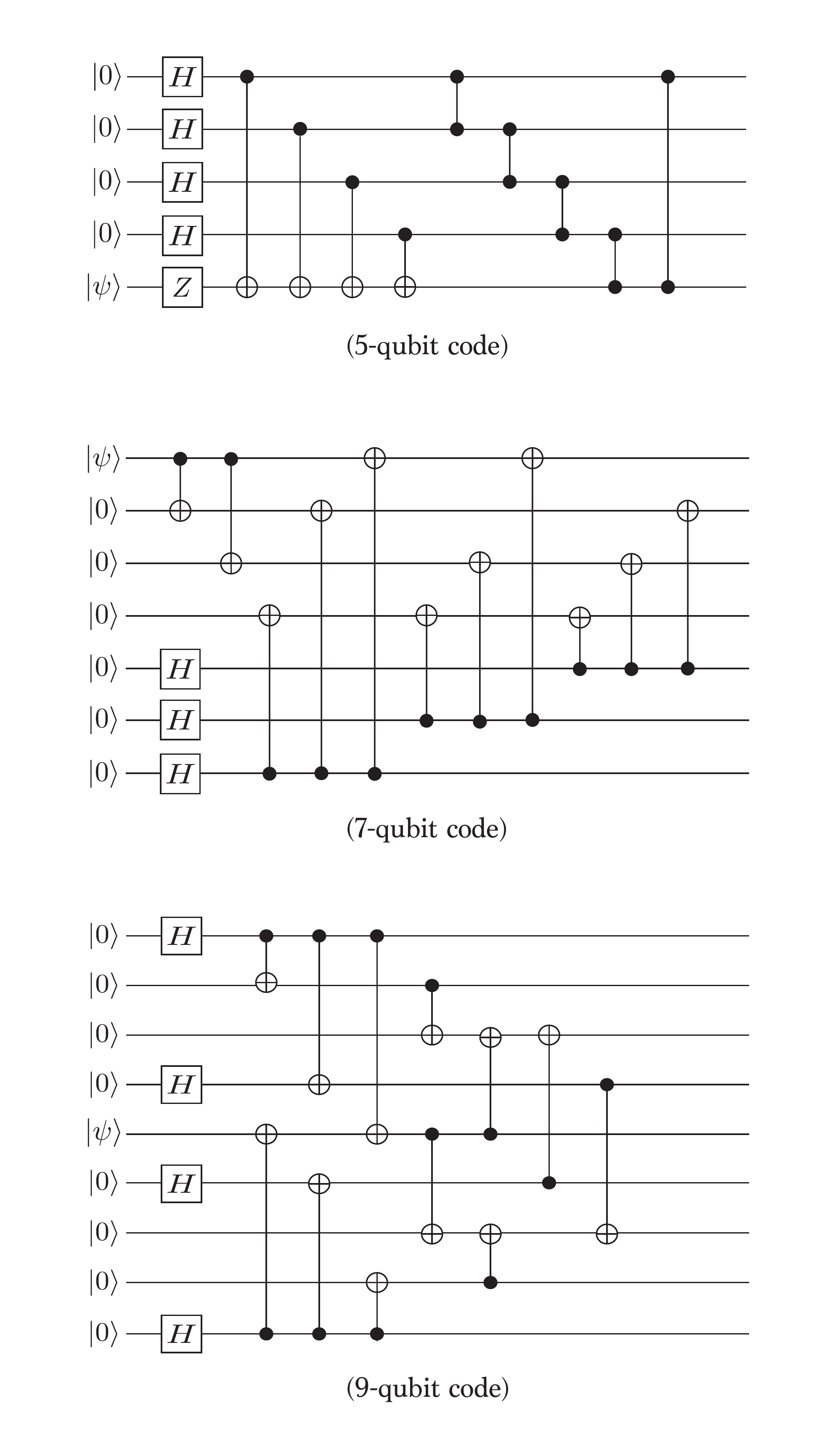}
\caption{{\bf General encoding circuits suitable for the five-qubit, Steane, and nine-qubit codes.} In all cases the physical state of $\psi$ is encoded into the logical state $|\psi\rangle_L$ }
\label{fig:circuitsRandom}
\end{figure}

In Figure~\ref{fig:circuitsRandom} we show the encoding circuits which we employ when Alice (taken to be ideal) encodes the physical qubit $\ket{\psi}$ which she has chosen to place into the memory. The encoding circuits come from Refs.~\cite{de2016universal},~\cite{divincenzo2007effective} and ~\cite{li2015magic}, respectively. Because Alice is perfect, there is no need for fault tolerance in these encoders. Bob employs the inverse of these encoders as a step in his analysis, see Table~\ref{table:basic_protocol}.

\begin{figure}
\centering
\includegraphics[width=0.9\linewidth]{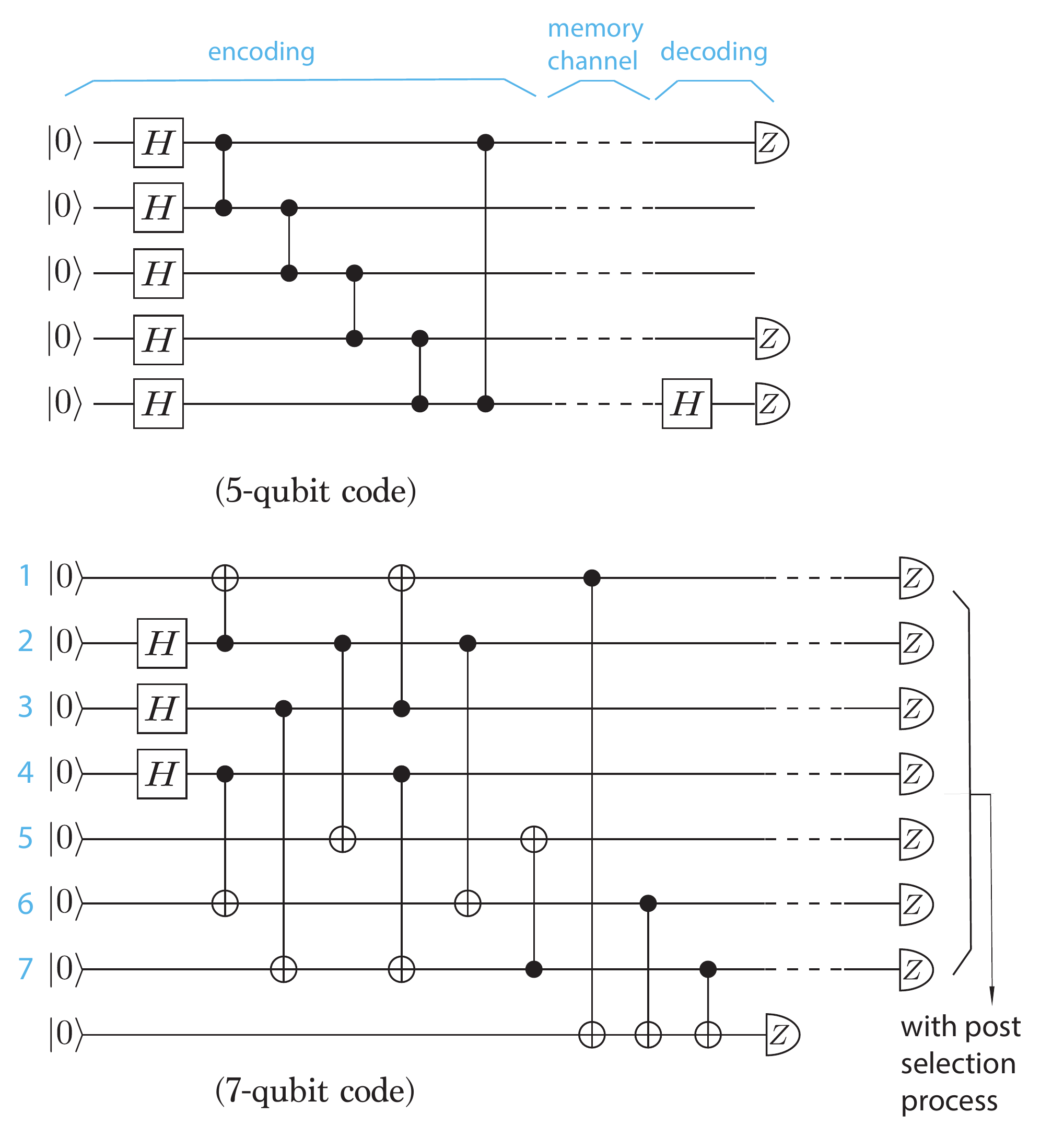}
\caption{{\bf Encoding circuits for imperfect encoding procedures with the five- and seven-qubit codes.} For the five-qubit code shown in the upper panel, the encoded state $|-\rangle_L$ is prepared in a non-fault-tolerant fashion, and Bob subsequently identifies the received state by measuring three of the received qubits and computing their parity (again, a non-fault-tolerant process). For the seven-qubit code shown in the lower panel, physical qubits are encoded into $|0\rangle_L$ using additional qubit for detection of errors: if returns 1, Alice restarts the encoding until it returns 0. Such method reduces propagation of some errors in a noisy encoding process. Bob is also fault tolerant: he measures all 7 qubits, and may opt to flip one of the outcomes if it is necessary to do so in order to produce a legitimate outcome; the parity of subsets 4,5,6,7; 1,3,5,7; 2,3,6,7 should all be the same as to allow him to guess between  $|0\rangle_L$ and  $|1\rangle_L$.}
\label{fig:circuitsZero}
\end{figure}

In order to experimentally investigate the integrity of a memory channel, we must use circuits for Alice and Bob that are as compact as possible and, as a strong preference, fault tolerant. Fortunately we need not consider general encode/decode circuits since (for a broad class of noise models) we know that the worst-case choice of state for Alice to transmit will be a Pauli basis state. Thus it is such states that we need to Alice to prepare and Bob to differentiate. A suitable compact, fault tolerant encoding circuit for the Steane code is shown in Fig.~\ref{fig:circuitsZero}(b) which is adopted from Ref.~\cite{goto2016minimizing}. An equally compact, but non fault tolerant encoding circuit for the five-qubit code is shown in Fig.~\ref{fig:circuitsZero}(b). For both the seven-qubit and the five-qubit cases, our Bob now simply measures all the qubits; however importantly for the seven-qubit case he can perform classical error correction on the measurement results making his inference process fault tolerant.

\subsection{Igor's circuits}

\begin{figure}
\centering
\includegraphics[width=1\linewidth]{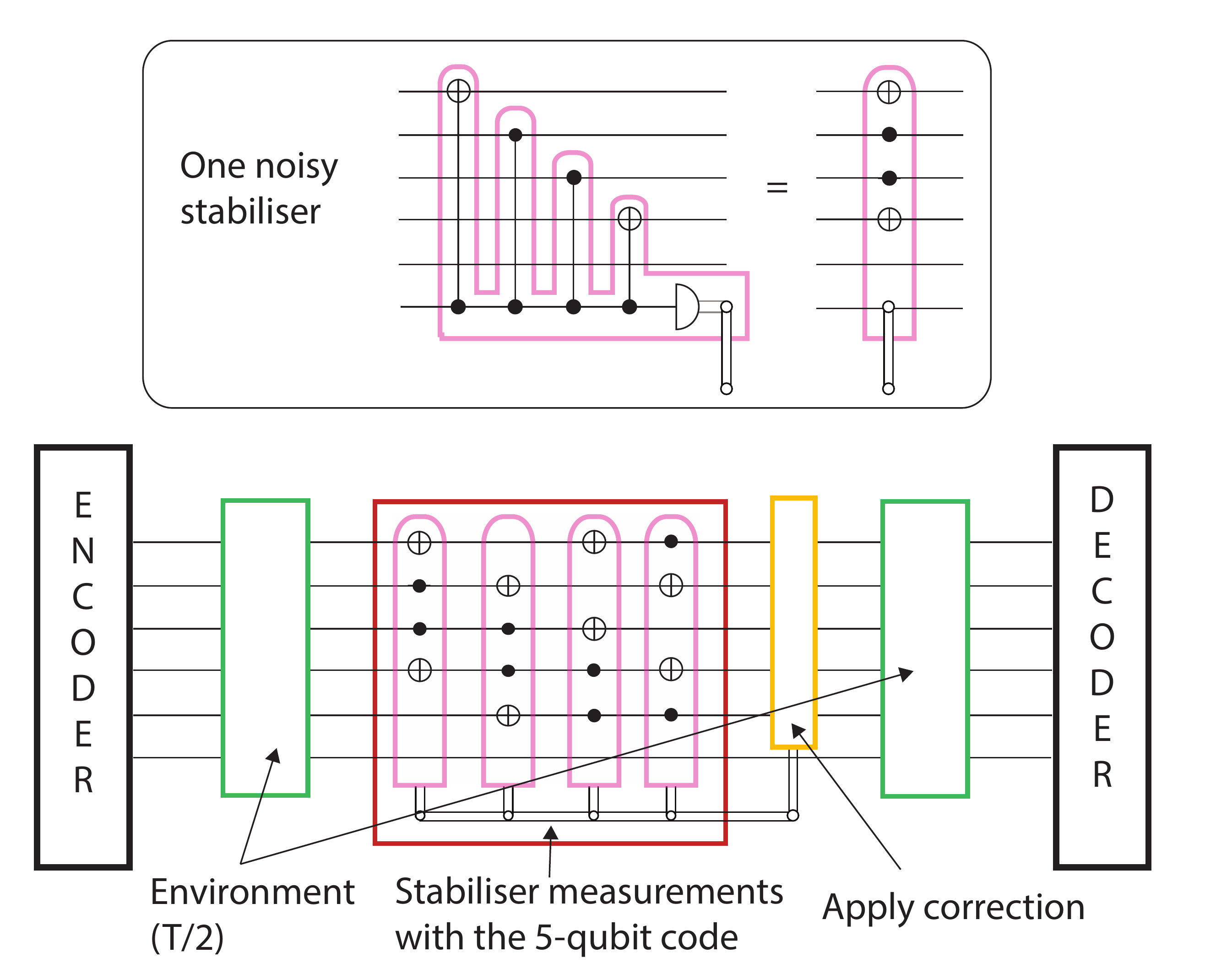}
\caption{{\bf Diagram of one whole cycle of Alice-Igor-Bob scenario with the five-qubit code.} Firstly five physical qubits are encoded into the logical state, then the logical qubit is subjected to environmental noise for a time period of $T/2$, followed by a cycle of stabilizer measurements and error correction, and then the logical qubit is again subjected to environmental noise for $T/2$. Lastly the logical qubit is decoded and measured. }
\label{fig:5qubitScheme}
\end{figure}

\begin{figure*}
\centering
\includegraphics[width=1\linewidth]{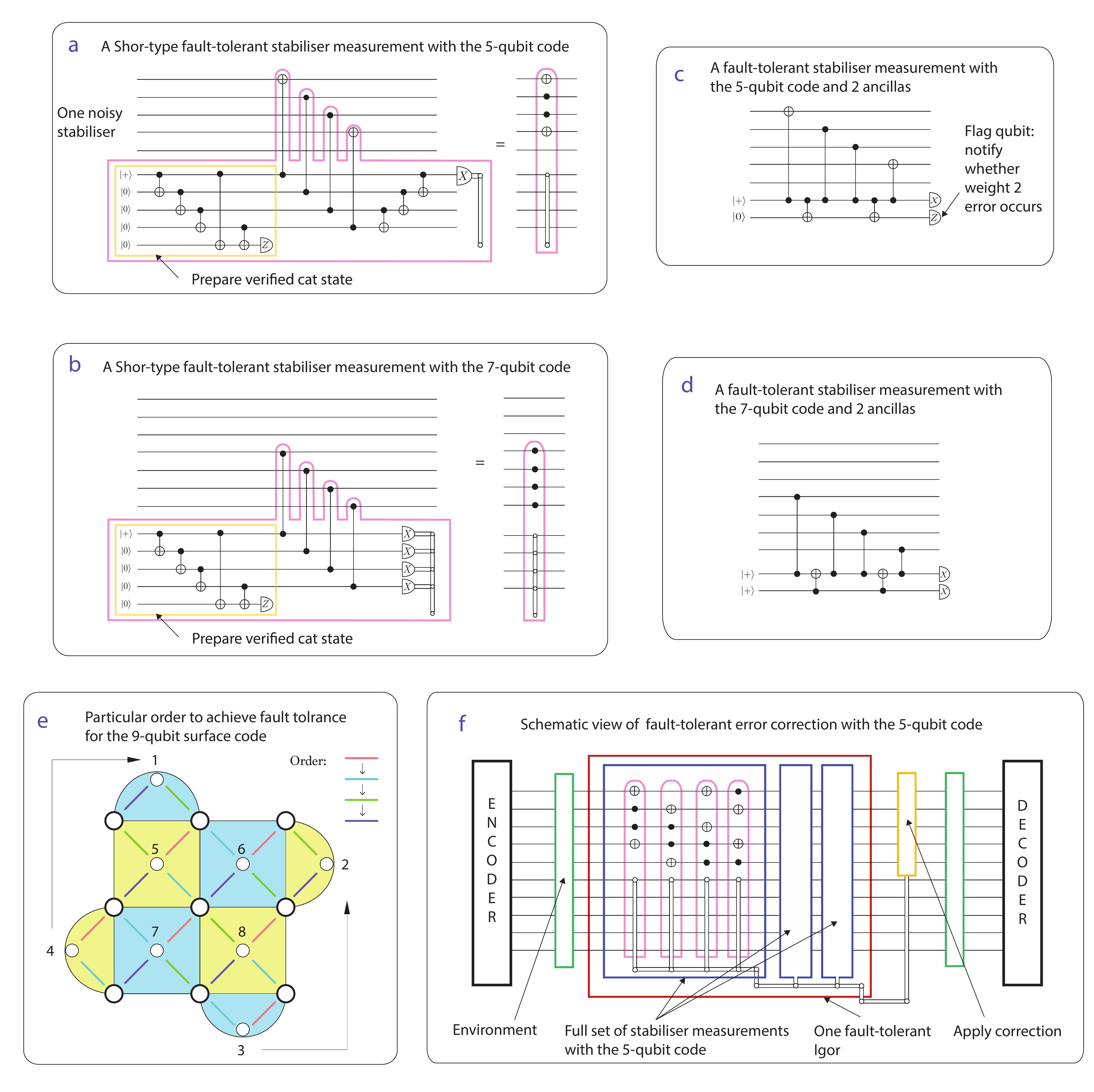}
\caption{{\bf Circuits and diagrams for fault tolerant error correction. }
(a) a round of fault tolerant stabilizer measurement with the five-qubit code. Here we use the Shor's method, with which four ancilla qubits are entangled into the cat state with some error probability and an additional qubit is used to verify the cat state is successfully prepared. After the stabilizer measurement, the ancillas are decoded, followed by measurement in $x$-basis. 
(b) a round of fault tolerant stabilizer checks with the seven-qubit Steane code, again using Shor's method. Since each stabilizer check detects either phase or bit flips, results can be obtained by checking the parity of measurement results of all four ancillas without decoding. 
(c) the circuit to achieve fault tolerant correction of the five-qubit code with only two ancillas. The first ancilla is used for stabilizer measurement, while the other one acts as the flag qubit: it returns -1 once any weight 2 errors occurs, and all such errors render a unique error syndrome thus can be corrected. 
(d) the same approach as in (c), but for for the seven-qubit Steane code. 
(e) stabilizer measurements of ancillas following a particular order to achieve fault tolerance with the rotated nine-qubit surface code. The large circles stand for the data qubits and the small circles are ancillas. The stabilizer measurement should follow the order denoted by the colour orange, blue, green and finally purple. Since the ancilla labelled 3 can physically act as the ancilla labelled 2 after finishing the measurement in the blue half-circle and that also works for ancilla 4, which can act as ancilla 1 after the measurement in the yellow half-circle, in total six ancillas are required to demonstrate fault tolerance.
(f) schematic view of the whole cycle of Shor-type fault-tolerant Alice-Igor-Bob scenario with the five-qubit code ((a) in this figure). The same procedure also works for all the others described above. Three rounds of a full set of stabilizer measurements are required to obtain the correct error syndrome as to avoid artificially introducing new errors through error correction based on the wrong syndrome.  
 }
\label{fig:FTScheme}
\end{figure*}

Figure~\ref{fig:5qubitScheme} shows the entire Alice-Igor-Bob process. In this figure, the memory channel employs the five-qubit code and Igor's error correction is not fault tolerant. Consequently the overall circuit is one of the more simple examples; but cases where we employ the Steane code or the nine-qubit code are analogous, as are cases where we opt to make Igor's process fault tolerant. The specific sub-circuits for these cases are shown in Figure~\ref{fig:FTScheme}. 

In our simulations (e.g. Fig.~\ref{fig:IgorError}) we considered more than one type of fault tolerance. The most common method to avoid weight-2 errors is to encode four ancilla qubits into a cat state, verified with additional qubit, and apply transversal CNOT gates within each stabiliser check, which may be known as Shor's method. Circuits in Figure~\ref{fig:FTScheme} (a) and (b) demonstrate this approach. A slight difference between these two diagrams exists, regarding measurement of the ancilla: for the five-qubit code, the encoded ancilla qubit needs to be decoded by applying the gates used in encoding in reverse before measuring the decoded physical qubit, while for the Steane code, since each stabiliser check detects only one type of error, we can simply measure all the four physical qubits in the corresponding basis and check the parity of the measurement results. 

The alternative fault-tolerant circuits with only two ancilla qubits are shown in Figure~\ref{fig:FTScheme} (c) and (d) for the five-qubit and seven-qubit codes, respectively. Here we are employing the ideas recently introduced in Ref.~\cite{chao2017quantum}. The first ancilla qubit acts the same as that in the non-fault-tolerant circuit, and the second ancilla qubit acts as the flag qubit: once any weight-2 error occurs, the measurement of it will turn from $0$ to $1$. For both the five-qubit and seven-qubit codes, each weight-2 error corresponds to a unique error syndrome if applying a set of normal stabiliser checks, thus we can detect any weight-2 error by measurement of the flag qubit and correct by mapping the stabiliser measurement results with the unique error syndrome. 

The nine-qubit code has the unusual and desirable property that the techniques described above, involving multiple ancillas, are not needed for fault tolerance. As shown in Fig.~\ref{fig:FTScheme}(e), weight-2 errors can be avoided simply by taking care to measure the stabilisers in a certain order (as has been discussed in Ref.~\cite{svore2014} and Ref.~\cite{wootton2016noise}). Since only one round of stabiliser checks is to be preformed, fewer gates compensate the cost of six ancilla qubits required. 

The full diagram for evaluating the memory with fault-tolerant error correction is shown in Figure~\ref{fig:FTScheme} (f), where we take the Shor-type five-qubit code (Figure~\ref{fig:FTScheme} (a)) as an example -- analogous circuits apply for the other cases. Compared with the non-fault-tolerant error correction as shown in Figure~\ref{fig:5qubitScheme}, three rounds of stabiliser measurements are required in order to avoid additional errors introduced by error correction based on wrong error syndromes. 

\section{Comparison with a more powerful Bob}
\label{uberBob}
\begin{figure}
\centering
\includegraphics[width=1.05\linewidth]{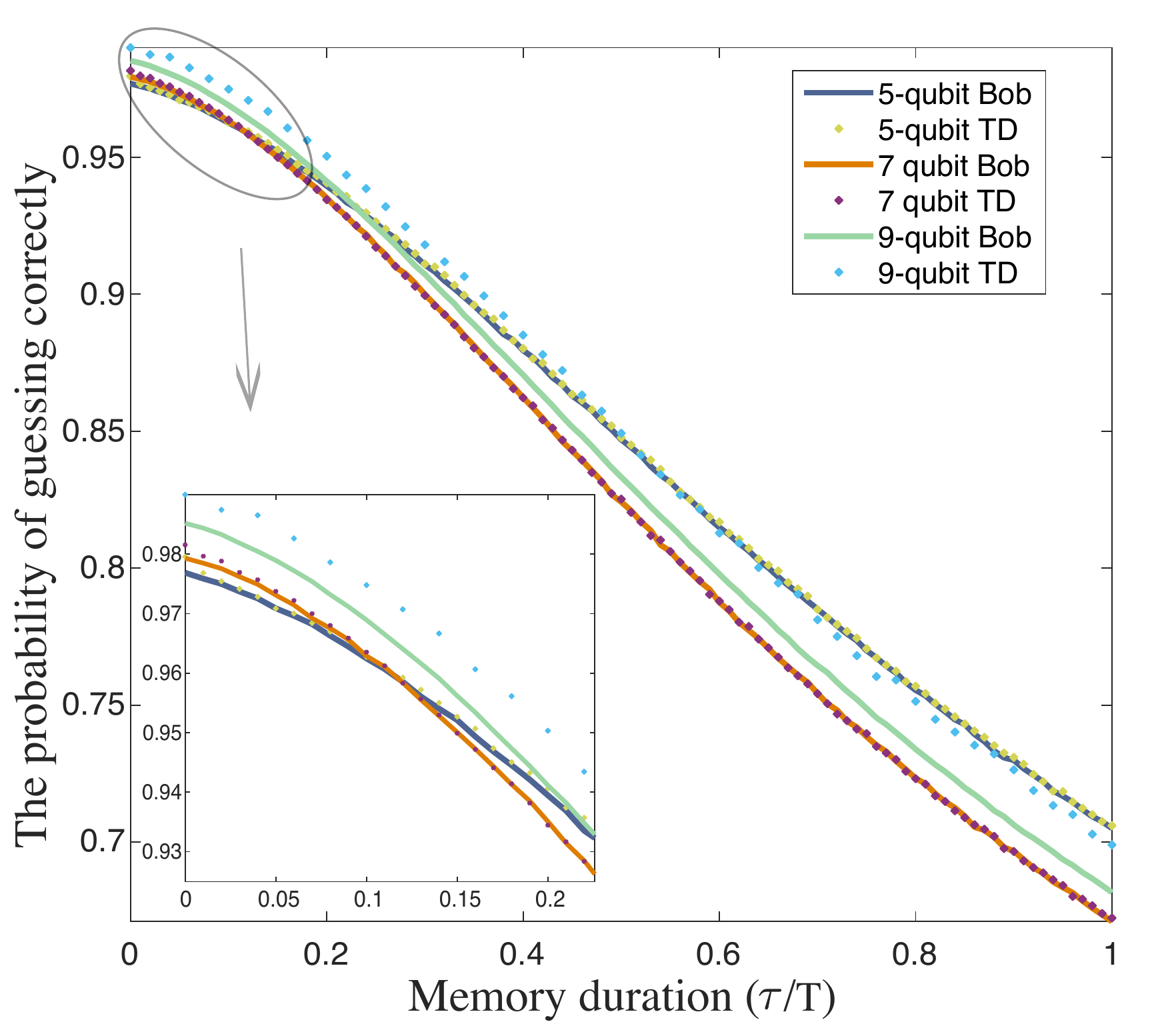}
\caption{{\bf Effect of a `more powerful' Bob.} The solid lines here correspond to the performance of Bob as we have specified him within our definition of integrity. The vertical axis here is Bob's probability of making a successful guess, and the solid lines correspond to memory channels with a single round of Igor's error correction with error rate $0.5\%$. The dotted lines are the performance of a more powerful Bob as described in the text; the dotted and solid lines are essential identical except for the nine-qubit code.}
\label{fig:TDvBob}
\end{figure}

The integrity measure contains within its definition the notion that the agent Bob, who receives the memory state at the end of its duration, will perform a round of (perfect) error correction as the first step of his analysis. This ensures that we make good contact with existing concepts such as the fidelity of a logical qubit, or the logical error rate -- at least where those latter concepts have clear meanings. 

However it is an interesting exercise to to make a comparison between Bob's ability to correctly guess the received state, as captured by the integrity, versus Bob's performance if he were given carte blanche to make his guess by performing {\it any} physically allowed process on all of the encoded physical qubits. The performance of such a Bob would correspond to the trace distance 
\[
   p_B=\tfrac{1}{2}+\tfrac{1}{2}D\bigl(\ \rho_n^\prime,\rho_n\bigr), \;
\]
where $\rho_n$ is the encoded state prepared by Alice, and $\rho_n^\prime$ is the corresponding state subsequently received by Bob. In Figure~\ref{fig:TDvBob} we show a comparison between the performance of this more powerful Bob, and the Bob as we have defined for the integrity measure.  For both the five-qubit code and  the Steane code there is a negligible difference when Bob is given this extra freedom. For the nine-qubit code there is a small difference. This indicates that the error correction process itself is not quite optimal: some measurements differing from the canonical nine-qubit code stabiliser measurements would permit a superior guess, however such measurements might be very non-trivial to implement (generally the basis states can be entangled).

In order to achieve this slightly higher level of performance, Bob would require not only the freedom to make any measurements he sees fit on the received $n$ encoded qubits, but (crucially) also a complete understanding of the noise processes in the memory channel. In short, he would require an accurate theoretical description of the memory channel itself, so that he can derive both $\Phi(\psi)$ and $\Phi(\psi_\perp)$ once the two options for the original qubit, $\psi$ and $\psi_\perp$, are revealed to him. Only then can he determine what measurements to make in order to achieve maximum probability of a distinguishing between them. 

For these reasons we opt to constrain Bob as described in the main paper. Doing so gives us a more `operational' meaning to integrity, and allows us to make direct links to other related concepts in the field.

\section{Case where fault tolerance is beneficial}

\begin{figure}
\centering
\includegraphics[width=1\linewidth]{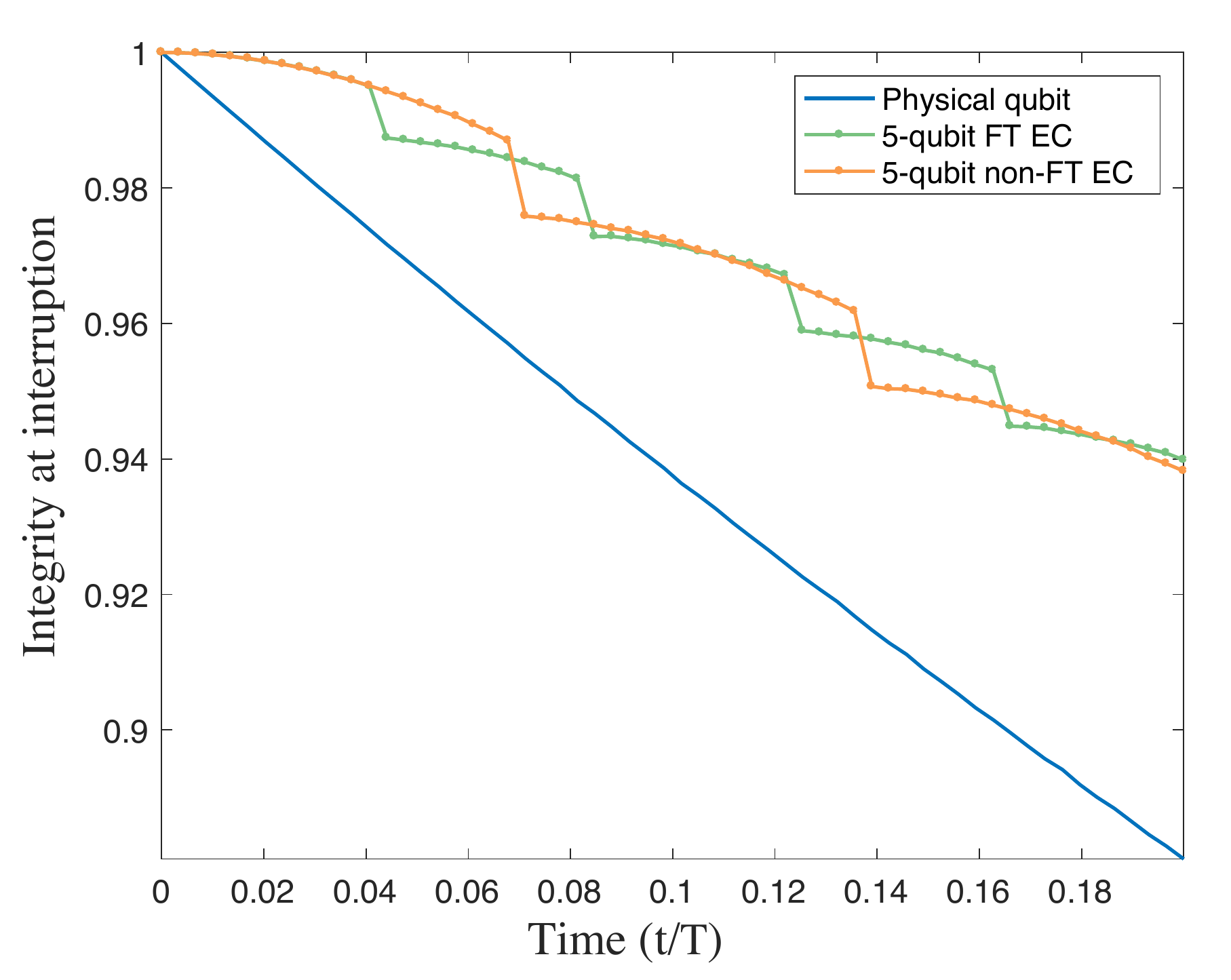}
\caption{{\bf Comparing memories employing FT verus a non-FT error correction.} In this figure we plot the `integrity at interruption' to look inside a memory process, as discussed earlier for Figs.~\ref{fig:multi-1} and~\ref{fig:multi}. We compare three memory channels all of the same duration $\tau=0.2\,T$. The blue line is our standard reference, the single-qubit memory. The other two are based on the five-qubit code, with the error rate of all the gates involved in the error correction process to be 0.1\%. The data shown in orange are for the memory channel protected by Igor using a non-FT error correction, and the optimal number of such cycles is 2. The data shown in green is for a more sophisticated Igor using the fault tolerant circuit shown in Fig.~\ref{fig:FTScheme}(c). It is interesting to note that the optimal number of error correction cycles is now $4$. However the overall performance is near-identical (i.e. lines are very close on the far right). \label{fig:interruptionFT}}
\end{figure}

In the main text, Fig.~\ref{fig:multiigor}(b) shows the performance of a high quality memory channel using the five-qubit code and performing $n$ cycles of error correction, equispaced over the duration, with a gate error rate of $0.1\%$. As noted in that figure caption and the associated main text, with this level of fidelity we find that the memory channel meets the most demanding of our milestones, {\it M4: Strictly superior encoded memory}. In the case analysed in the main paper Igor used a non-fault tolerant error correction cycle; however from the earlier Fig.~\ref{fig:IgorError} one would expect that a fault tolerant Igor using the (recently proposed) 2-ancilla technique might lead to an {\it even} more highly performing memory channel.

In fact we have found that the memory using fault-tolerant correction is indeed superior, albeit just slightly. An interesting point is that the optimal number of error correction cycles is {\it higher}, for a given channel duration, when one employs fault tolerant correction versus the naive circuit. This makes intuitive sense: when gate errors are as low as $0.1\%$ the fault tolerant error correction circuits work well and introduce less noise than the naive circuits (c.f. Fig.~\ref{fig:IgorError}), so that we will see smaller step-like deteriorations in the quantity we call `integrity at interruption' implying that they can be used more frequently. This is shown in Fig.~\ref{fig:interruptionFT} where we contrast a fault-tolerant and non-fault-tolerant channels of duration $\tau=0.2\,T$.

\section{Significance of imposing a minimum }

\begin{figure}
\centering
\includegraphics[width=1\linewidth]{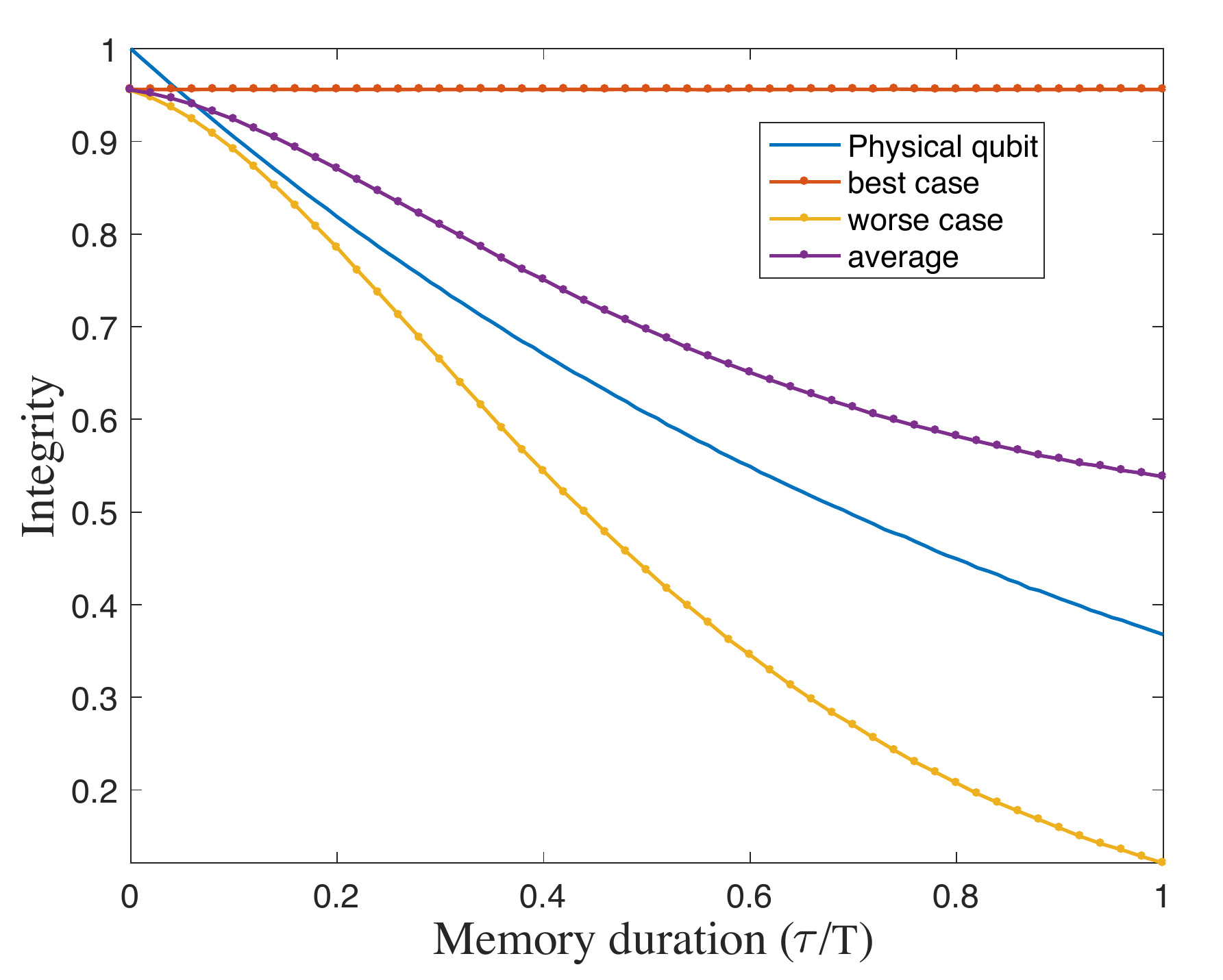}
\caption{{\bf Integrity in a pure dephasing environment.} Plot shows the integrity for a memory channel using a the seven-qubit Steane code, and and single round of Igor's error correction procedure with a gate error rate of $0.5\%$. Whereas all other plots in this paper correspond to a pure depolarising environment, here we have a pure dephasing environment. Consequently Bob's ability to guess the nature of the received state depends strongly on Alice's choice of which qubit to send: If she sends a $z$-basis eigenstate then Bob's success is certain. Note that while the $y$-axis is labelled `Integrity' only the blue and yellow lines conform to the definition, Eqn.~(\ref{eqn:integrityShortDef}).  }
\label{fig:worstcase}
\end{figure}

In all simulations previously described in this report the environmental decoherence was purely depolarising. Consequently the environment has no preferred basis, and one finds that Bob's probability of successfully guessing the nature of the state selected by Alice does not vary according to her choice. Thus the minimum appearing in the definition of integrity, Eqn.~\ref{eqn:integrityShortDef}, is redundant in the sense that the minimum and maximum are the same. In order to show that this will not generally be true, and that therefore it is indeed necessary to specify the minimum, we need only switch from a pure depolarising enviroment to a pure dephasing environment. 

The results of such simulations are shown in Fig.~\ref{fig:worstcase} which shows a Steane-code protected memory as in earlier plots (c.f. Fig.~\ref{fig:sevennoisyAB}) but now with all environmental noise being pure dephasing. The interesting point is that now Bob's ability to guess the original encoded state varies dramatically with Alice's choice of initial state. If she chooses either $\ket{0}$ or $\ket{1}$ then the logical qubits are in fact immune to phase noise, so that Bob's performance impaired only by the noise introduced by Igor -- the corresponding line (red) is therefore flat i.e. not a function of the memory duration. In contrast Bob's `worst case' performance is obtained when Alice's choice for the encoded state is $\ket{+}$ or $\ket{-}$ as shown by the yellow line, and it is this that would define the integrity of memory channel.

In the following section we explain for many common environmental noise models the `worst case' will be found among the Pauli eigenstates.

\section{When does it suffice to prepare Pauli eigenstates?}\label{appendix:whenIsPauliEnough}

In the main text and in the preceding appendix we noted that Alice's choice of state to encode can influence Bob's performance when he guesses the nature of the received state. Therefore integrity is defined from the worst case performance.  In the main text we noted that when indeed this occurs, we will often find that the worst case corresponds to Alice choosing a Pauli eigenstate. Here we explain that this is typical for a broad range of error models. In the following, when we refer to `weak' noise this is in the sense that the error probability is $\le 0.5$, which is in general the region of interest where the milestones M1-M4 can be met.

Recall that we are assessing single-qubit memories, represented by a channel $\Phi$ on single-qubit states, in the presence of realistic noise using experimentally viable methods.
We proceed by describing a general class of noise channels with noise dominated by incoherent Pauli errors, and describe conditions under which we can reduce the testing of single-qubit memories subject to such noise to testing of Pauli eigenstates.

We write $\mathbf P = \{\sigma^{(0)}\!, \sigma^{(1)}\!, \sigma^{(2)}\!, \sigma^{(3)}\}$ for the set of single-qubit Pauli operators (we sometimes write $\idop = \sigma^{(0)}$, $X = \sigma^{(1)}$, $Y = \sigma^{(2)}$, and $Z = \sigma^{(3)}$).
Let $\mathbf P^M$ denote the set of $M$-fold tensor products of Pauli operators $\sigma^{(a_1)} \ox \cdots \ox \sigma^{(a_M)}$.
A {\it Pauli channel} is a channel on $M \ge 1$ qubits whose Kraus operators are each proportional to an element of $\mathbf P^M$, representing random Pauli operators acting on those qubits according to some distribution.
We may denote such a channel by
\begin{equation}
  \Phi(\rho)  \,=\! \sum_{\tau \in \mathbf P^M} \!\! p_\tau \; \tau \rho \:\!\tau\herm.  
\end{equation}
(All such channels are unital, i.e.~$\Phi(\tfrac{1}{d}\idop) = \tfrac{1}{d}\idop$.)

A {\it weak Pauli channel} is such a channel in which $p_{\idop \ox \cdots \ox \idop} \ge \tfrac{1}{2}$.
We can experimentally estimate the integrity of a weak Pauli channel $\Phi$ on a single qubit, as follows.
First note that we may simplify the formula of memory integrity from Eqn.~\eqref{eqn:integrityShortDef} by noting that
\begin{equation}
\begin{aligned}[b]
    \mathcal R(\Phi)
  \,&=
    \min_{\psi_0 \perp \psi_1}\,
      \tfrac{1}{2} \bigl\lVert\Phi(\psi_1) - \Phi(\psi_0) \bigr\rVert_\tr
  \\[1ex]&=
    \min_{\psi}\;
      \bigl\lVert\Phi(\psi) - \Phi(\tfrac{1}{2}\idop) \bigr\rVert_\tr 
      \;.
\end{aligned}
\end{equation}
For such channels we have $\Phi(\sigma^{(j)}) = \alpha_j \sigma^{(j)}$ for some $\alpha_j \ge 0$: in particular, as these channels are unital, $\alpha_0 = 1$.
Then for any single-qubit state $\rho = \tfrac{1}{2}\bigl[\idop + r_1\sigma^{(1)} + r_2 \sigma^{(2)} + r_3 \sigma^{(3)}\bigr]$, we have
\begin{equation}{}\!\!\!\!
\begin{aligned}[b]
    \Bigl\lVert \Phi(\rho) - \Phi(\tfrac{1}{2}\idop) \Bigr\rVert_\tr
  \!&=\;
    \Biggl\lVert \sum_{j=1}^3 \tfrac{1}{2} r_j\alpha_j \sigma^{(j)} \Biggr\rVert_\tr
  \!=\;
    \sum_{j=1}^3 r_j^2 \alpha_j^2 .
\end{aligned}
\!\!\!\!\!\!\!
\end{equation}
This is a convex combination of the scalars $\alpha_j^2$, which is minimised by setting $r_j^2 = 1$ for the smallest coefficient $\alpha_j$ and $r_j = 0$ otherwise.
Thus
\begin{equation}{}
\label{eqn:memoryQualityRandomPauli}
\mspace{-15mu}
  \mathcal R(\Phi) \!\:=\!\: \min_j \alpha_j^2 \!\:=\!\: \min_j \,\mathcal D\Bigl(\!\!\:\Phi(\varphi_+^{(j)}), \Phi(\varphi_-^{(j)}) \!\!\:\Bigr),
  \!\!
\end{equation}
where $\varphi_\pm^{(j)}$ are the $\pm1$-eigenstates of the respective Pauli operator $\smash{\sigma\sur{j}}$.

A {\it weak i.i.d.\ Pauli channel} is such a channel which consists of a tensor product $\Phi_1 \ox \Phi_1 \ox \cdots \ox \Phi_1$ of identical channels.
We are more generally interested in maps $\Phi = \mathsf D \circ \mathsf N \circ \mathsf E$ which consist of an ideal encoder $\mathsf E$ for a stabiliser code (encoding one qubit into $M$ qubits), a noise process $\mathsf N$ weak i.i.d.\ Pauli channel on $M$ qubits, and an ideal decoder $\mathsf D$ which performs one round of correction decodes the $M$ qubit state again to a single-qubit state.
It is not difficult to show that $\Phi$ will be a weak Pauli channel when $\mathsf N$ is a weak i.i.d\ Pauli channel, in which case the worst-case performance will be achieved by Pauli eigenstates in this case as well.

This motivates the following procedure to experimentally assess the quality of an isolated quantum memory on a weak Pauli channel: prepare a state $\smash{\varphi\sur{j}_\pm} = \ket{\phi}\bra{\phi}$, apply $\Phi$ to it, and test the probability with which we obtain the outcome $\ket{\phi}\bra{\phi}$ when a $\sigma^{(j)}$ measurement is performed on it.
Performing the above many times for each Pauli operator $\sigma^{(j)}$, we may determine with some level of confidence for which operator $\sigma\sur{j}$ this fails most often.
This determines the pair of orthogonal states which $\Phi$ does the poorest job at keeping distinguishable; using Eqn.~\eqref{eqn:integrityRelatesToP}, we may then compute $\mathcal R(\Phi)$.

\end{document}